\documentclass[review]{elsarticle}

\usepackage{lineno,hyperref}
\usepackage{makecell}

\modulolinenumbers[5]

\journal{}

\usepackage{color}
\usepackage[normalem]{ulem}

\begin{document}

\begin{frontmatter}

\title{Strategies for particle resampling in PIC simulations}
\author[IAP]{A.~Muraviev}
\author[IAP]{A.~Bashinov}
\author[IAP]{E.~Efimenko}
\author[UNN]{V.~Volokitin}
\author[UNN]{I.~Meyerov}
\author[GU]{A.~Gonoskov}

\address[IAP]{Institute of Applied Physics, Russian Academy of Sciences, Nizhny Novgorod 603950, Russia}
\address[UNN]{Lobachevsky State University of Nizhni Novgorod, Nizhny Novgorod 603950, Russia}
\address[GU]{Department of Physics, University of Gothenburg, SE-41296 Gothenburg, Sweden}

\date{\today}

\begin{abstract}
In particle-in-cell simulations, excessive or even unfeasible computational demands can be caused by the growth of the number of particles in the course of prolific ionization or cascaded pair production due to the effects of quantum electrodynamics. Here we discuss how one can organize a dynamic rearrangement of the ensemble to reduce the number of macroparticles, while maintaining acceptable sampling of an arbitrary particle distribution. The approaches of merging and thinning as well as their variants are discussed and the aspects of use are considered.

\end{abstract}


\end{frontmatter}

\section{Introduction}

In particle-in-cell (PIC) simulations and some other statistical computations the use of so-called macroparticles can be supplemented by the process of adding new macroparticles. For example, in PIC plasma simulations this can be done to account for continuous ionization of matter \cite{bruhwiler.pop.2003, chen.jcp.2013} or for electron-positron pair production due to the effects of quantum electrodynamics (QED)\cite{nerush.prl.2011, elkina.prstab.2011, sokolov.pop.2011, ridgers.jcp.2014, gonoskov.pre.2015, chang.2017}. When such particle sources become prolific the number of macroparticles can grow significantly. This can slow down the simulation and/or exhaust the memory available for the allocation of macroparticles’ data. The natural solution is to resample the modeled distribution using a smaller number of macroparticles with an increased weight of their contribution. This procedure, referred to as down-sampling, can be repeatedly applied to combat the growth of the ensemble of macroparticles or, alternatively, to reduce the computational expenses for highly populated regions of phase space, where the representation has become excessive with time. In the latter case, the released resources can be used to reduce the computational noise in underpopulated regions by introducing more macroparticles for better sampling, i.e. performing the so-called up-sampling. 

The implementation of down-sampling has been considered by several authors and a number of methods have been proposed. One can distinguish three main approaches. According to the first approach, referred to as \textit{merging} or \textit{coalescing}, one (or two) macroparticles is introduced to replace a subset of macroparticles that are close in the phase space. According to the second approach, referred to as \textit{thinning}, we do not introduce new macroparticles, but remove one or several macroparticles and redistribute their weight among the others, either globally or locally within the given subset. Finally, one can totally replace the given subset of macroparticles with a new subset of appropriately introduced new macroparticles. This is the third approach, which is referred to as \textit{complete resampling} in this article. In all cases, for the selected subset of closely located macroparticles the procedure can potentially lead to the change of intrinsically conserved quantities, such as the total mass (weight), charge, energy or momentum. In addition, the procedure can potentially lead to the sudden change of grid values for the charge and current density, leading to artificial noise, heating/cooling or other systematic effects. That is why the possibility of preserving such quantities has been considered with special care by many authors.

Within the approach of merging Lapenta and Brackbill proposed a method for coalescing two particles into one so that the charge assignment to the grid nodes is preserved \cite{lapenta.jcp.1994}. Merging several macroparticles of dense clusters selected with the use of the Voronoi algorithm has been suggested by Luu et al. \cite{luu.cpc.2015}. To conserve both energy and momentum Vranic et al. proposed to perform merging into a pair of particles with the appropriately chosen momenta \cite{vranic.cpc.2015} (this idea has been earlier considered in Refs.~\cite{rjasanow.jcp.1996, rjasanow.jcp.1998}). This can be arranged using the selection of highly populated volumes in the phase space. A similar method with some modifications has been used in Ref.~\cite{chang.2017}.

The approach of thinning provides various options, including rather straightforward ones.  For example, Timokhin developed a procedure that repeatedly selects a random particle, deletes it and uniformly redistributes its weight among the others of the same kind \cite{timokhin.mnras.2010}. Nerush et al. used a similar global thinning but with the redistribution of mass, charge and energy \cite{nerush.prl.2011}. Although this procedure preserves the mentioned quantities globally, it implies their stochastic local variations at the instant of resampling. One way of preventing such variations is to restrict the redistribution to a dense cluster or a highly populated volume of the phase space. A way to perform thinning with conservation of several arbitrary particle and grid quantities is proposed in Ref.~\cite{gonoskov.arxiv.2020}.

Within the approach of complete resampling, Lapenta and Brackbill proposed a method of replacing the macroparticles in a given cell with a new subset, preserving the contributions to the grid quantities and also maximizing the uniformity of the distribution of these quantities over the new subset \cite{lapenta.cpc.1995}. A way of doing such resampling with the conservation of grid values for charge and current density, as well as of the total energy of the resampled macroparticles, has been proposed by Assous et al. \cite{assous.jcp.2003} and further developed in Ref.~\cite{welch.jcp.2007}. Pfeiffer at al. proposed a statistical method that conserves momentum and energy \cite{pfeiffer.cpc.2015}. Faghihi et al. reported on the development of an algorithm for both down- and up-sampling that preserves any number of particle and grid quantities \cite{faghihi.arxiv.2017}.

Recently, the methods of down-sampling have become highly in demand for numerical studies of QED cascades that will be inherent for the upcoming experiments at the next generation high-intensity laser facilities \cite{ELI-NP, ELI-beamlines, CoReLS, SULF}. The numerical studies indicate that, apart from the drastic increase of the amount of particles by several orders of magnitude, the physics in strong laser fields includes a variety of new phenomena \cite{gonoskov.prl.2014,Muraviev.jetplet.2015,ji.prl.2014, gonoskov.prx.2017, tamburini.scirep.2017, vranic.scirep.2018, efimenko.scirep.2018, efimenko.pre.2019}. The absence of prior knowledge about the minimal scales of new phenomena raises a new difficulty for the implementation of down-sampling: merging macroparticles within dense clusters or volumes of predetermined scale may erase or affect smaller scale peculiarities that are essential for the modelled phenomena. Although the coordinates have the cell size as a natural limiting scale, the momentum does not have any natural resolution limit according to the PIC method. At the same time, narrowing the permitted momentum difference for merging in a given cell increases the density required for the selection of several particles to be possible.

In this article we consider how one can use the thinning approach or modify the merging procedure to combat the outlined difficulties. In addition, we consider the aspect of reducing the difference between the weights of particles as a way to increase the efficiency of sampling. For our study we develop, compare and analyze several methods that we release as open-source tools available within the hi-$\chi$ framework \cite{hi-chi}. The article is arranged as follows. In section 2 we introduce the principle of agnostic down-sampling for developing the thinning procedures that are applicable without prior knowledge about minimal scales attributed to the modelled process. In section 3 we describe several thinning methods that comply with this principle and also discuss how to modify the merging methods to improve their applicability in this context. In section 4 we perform an elementary comparison of the methods. In section 5, we analyze the use of methods on realistic problems. We conclude in section 6. 

\section{The principle of agnostic down-sampling}

\begin{figure}[t]
	\centering
	\includegraphics[width=0.5\columnwidth]{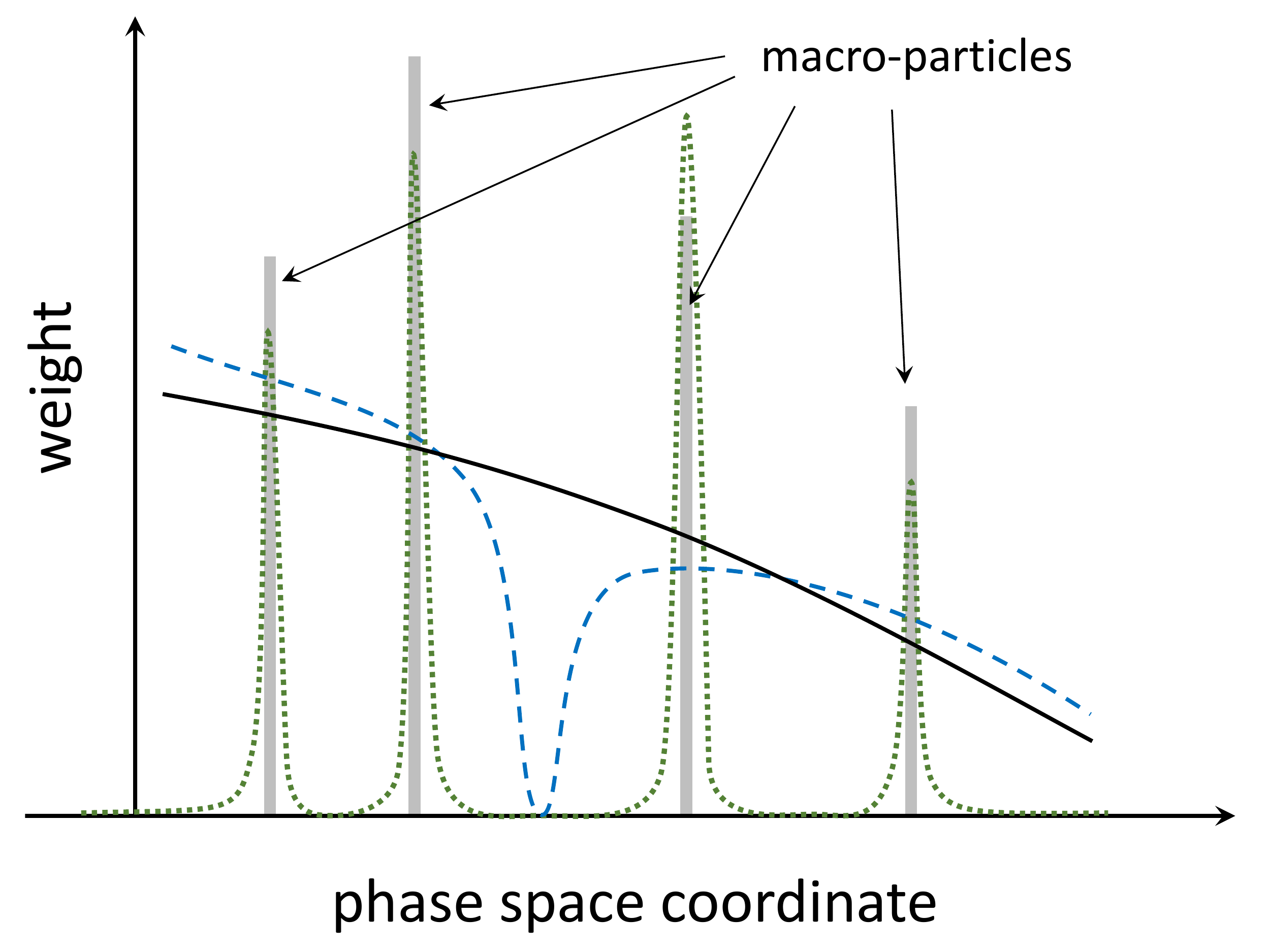}
	\caption{An illustrative clarification of the difficulty in arranging non-destructive merging without prior knowledge about the peculiarities in the particle distribution and their minimal scales. The merging of four macroparticles (grey rectangular peaks) can be non-destructive if they sample the distribution show with black solid curve. However it can affect the modelled processes if the sampled distribution has a more complex shape. The examples of such potential shapes are shown with dashed blue and dotted green curves. Color online.}
	\label{example}
\end{figure}

When several macroparticles are merged we assume that all these particles sample a uniform part of the particle distribution without any complex trends or peculiarities represented by some of them. For example, the macroparticles shown in Fig.~\ref{example} are assumed to sample the distribution shown with the solid curve, not the ones shown with the dashed or dotted curves. Although one can probably mitigate the risks of such misinterpretation by restricting the choice of particles in a cluster to sufficiently small phase space volumes, let us consider an alternative down-sampling methodology that is applicable without any prior knowledge and thus called agnostic here.

Firstly, note that simple merging into the average position of the selected macroparticles will likely reduce the variance of the peaks that are more narrow than the selection scale. To mitigate this one can introduce a probabilistic displacement according to the variance determined over the merged macroparticles. However, such a procedure will likely cause filling the gap in the distribution shown with the dashed curve in Fig.~\ref{example}. Looking at the distribution shown with the dotted curve we can see that the only way to not introduce any particles in any potentially empty phase space region is to use only the existing particles. In other words, we should try to restrict our action to the removal of one or several macroparticles in combination with the change of the weights of the other macroparticles.

Secondly, we should not affect any distribution functions. In order to do this, we can arrange a probabilistic procedure so that the chance of removing any particle in the given subset is compensated by the chance of increasing its weight. If the expectation value for the weights of each particle is exactly equal to its initial weight, all the possible distribution functions remain unchanged on average (see rigorous consideration below). This means that if some peculiarity is erased in one realization of resampling, it has a chance to be increased in a different one. In this way, the procedure only increases statistical variations but does not remove peculiarities at any scale.

We can now formulate the principle of down-sampling that is based on this idea. A down-sampling is called agnostic if it is restricted to the probabilistic change of weights so that (1) at least one macroparticle receives zero weight and can be removed and (2) the expectation value for the weight of each macroparticle is exactly equal to its initial weight.

If we have initially $n$ macroparticles and the weight of the $i$-th is $w_i$, the principle of agnostic down-sampling implies determining a number of outcomes so that the probability $p_k$ of choosing the $k$-th outcome provides
\begin{equation}
\left<\hat{w}_i\right> = \sum_{k \in \textrm{outcomes}}{w_i^k p_k} = w_i \:\:\: \textrm{for} \:\:\: i = 1, ... n,
\end{equation}
where $\left<\hat{w}_i\right>$ is the resulting weight averaged over the outcomes, and $w_i^k$ is the weight assigned to the $i$-th particle in the $k$-th outcome.

Let us demonstrate that any agnostic resampling has the property of preserving all the distribution functions on average, i.e. any distribution averaged over the outcomes of the probabilistic resampling procedure coincides exactly with that we had before the resampling. In order to show this, we consider an arbitrary distribution $G = \partial N / \partial g$, which can be numerically represented as a sequence of values $G_j$, representing the number of real particles, for which the quantity $g$ falls into the element $d_j$, divided by the volume of $d_j$:
\begin{equation}
G_j\left(\mathcal{A}\right) = \frac{1}{V\left(d_j\right)} \sum_{g\left({\bf x}_i, {\bf p}_i, {\bf \sigma}_i\right) \in d_j} w_i,
\end{equation}
where $g\left({\bf x}_i, {\bf p}_i, {\bf \sigma}_i\right)$ is the quantity (can be multidimensional) over which the distribution is computed, ${\bf \sigma}_i$ is a generalized vector of the particle's parameters other than coordinate and momentum (e.g. spin, polarization, etc.), $\mathcal{A}$ is the state of the ensemble, $d_j$ denotes the sub-regions used to discretize the space of $g$ values and $V\left(d_j\right)$ is the volume of $j$-th sub-region.  As one can see, in such a way we can define coordinate-, momentum-, energy-, angular- and other distributions on uniform as well as non-uniform grids defined by $d_j$. We can now formally compute the value $G_j$ averaged over outcomes of the resampling procedure that turns the state of the ensemble $\mathcal{A}$ into a new state $R(\mathcal{A})$:
\begin{equation}
\left<G_j\left(R(\mathcal{A})\right)\right>_R = \frac{1}{V\left(d_j\right)} \sum_{g\left({\bf x}_i, {\bf p}_i, {\bf \sigma}_i\right) \in d_j} \left<\hat{w}_i\right>_R = G_j\left(\mathcal{A}\right),
\end{equation}
where $\left<\hat{w}_i \right>_R$ is the $i$-th weight averaged over all realizations of resampling taking into account the corresponding probabilities. Here the first equality is provided by the fact that an agnostic resampling neither changes internal state $\left({\bf x}_i, {\bf p}_i, {\bf \sigma}_i\right)$ of nor adds new macroparticles, whereas the second equality follows from the fact that it preserves the weight on average, i.e. $\forall i: \left<\hat{w}_i \right>_R = w_i$. As one can see, we proved the statement without requiring any knowledge about (1) the quantity $g$, (2) the numerical intervals $d_j$ and (3) the distribution of macroparticles (either over the position in phase space or over the values of their weights). In this context the term \textit{agnostic} indicates that the procedure preserves all the distribution functions independently of the outlined entities. 

\section{Strategies of down-sampling}

In this section we propose a number of methods that comply with the principle of agnostic down-sampling. We start with the simplest methods and then describe more advanced ones and their potential benefits. 
We quantify the rate of resampling by a parameter $k$ being the target ratio of the number of macroparticles in the ensemble before resampling to their number after resampling.
Each method was given a shortened name in parenthesis for designation on graphics.

1. \textit{Simple thinning} (simple).

According to this method each macroparticle is either removed, with equal probability $p=1-k^{-1}$, or has its weight increased by a factor of $k$. This method is agnostic because $\left<\hat{w}_i\right> = 0 \cdot \left(1-k^{-1}\right) + kw_i \cdot k^{-1} = w_i$. The method does not strictly conserve any quantities such as total weight, energy or momentum, but conserves all quantities on average as any agnostic method. The method can be applied to subsets of any size, which makes it possible to apply it to very small volumes in phase space. This method is the easiest to implement and analyze theoretically. If the initial total number of macroparticles is $n$, the number of macroparticles remaining after resampling is approximately $n/k$.

2. \textit{Leveling thinning} (leveling).

In this method, we first calculate the average weight $\bar{w}$ of particles \textit{in a given cell}. Then, for all particles with weight $w_i < k \bar{w}$ the weight $k \bar{w}$ is assigned with probability $w_i/\left(k \bar{w}\right)$ and otherwise the particle is removed. This method is agnostic because $\left<\hat{w}_i\right> = k \bar{w} \cdot w_i/\left(k \bar{w}\right) + 0 \cdot (1 - w_i/\left(k \bar{w}\right)) = w_i$. It is clear that this procedure gets rid of macroparticles with weight below $k \bar{w}$. This may help to balance and optimize the distribution of computational resources. The method does not strictly conserve any quantities. The actual number of remaining macroparticles can be less or greater than $n/k$ depending on the initial weight distribution among macroparticles. If all macroparticles initially have the same weight, the method performs identically to the \textit{simple} method.

3. \textit{Global leveling thinning} (globalLev) is a modification of the previously described \textit{leveling} method. The method works similarly, except it accounts for particles in the entire computational region to compute the average weight $\bar{w}$. In case of parallel computations, given that each computational domain contains large number of cells, similar properties can be achieved by applying the method independently to each computational domain (no network transfers are needed in this case). We use the latter option in all computations in this paper.

4. \textit{Number-conservative thinning} (numberT).

In this method we select random macroparticles with probability proportional to their weight, i.e. $w_i/W$, where $W = \sum w_i$. We repeat this selection $m$ times and count the number of times $c_i$ we have chosen the $i$-th macroparticle. After that the particles that have not been selected even once ($c_i = 0$) are removed and the others are assigned with a new weight equal to $\hat{w}_i = c_i W/m$. It is clear that the $i$-th macroparticle will be selected $\left< c_i \right> = m w_i / W$ times on average and thus the mathematical expectation of the change in the macroparticle's weight is zero: $\left< \hat{w}_i\right> = \left( \left<c_i\right> W/m \right) = w_i$, i.e. the method is agnostic. This procedure strictly conserves the total weight of macroparticles in a cell: $\hat{W} = \sum{\hat{w}_i} = \sum{c_i W / m} = W$. In addition, it favors the removal of macroparticles with small weight. This may also contribute to the efficiency of sampling. The number of macroparticles $\hat{n}$ after this procedure is probabilistic but obviously cannot exceed $m$. {The average number of remaining macroparticles is given by the expression  
\begin{equation}
\hat{n} = \sum_{i = 1}^n{\left(1 - \left(1 - \frac{w_i}{W}\right)^m\right)}.
\end{equation}}
If we assume that in our distribution macroparticles have similar weights we can estimate $\hat{n} \approx n \left(1 - \left(1 - n^{-1}\right)^m \right)$. Assuming also that $n$ is large, we can estimate that $\hat{n} = n/k$ is achieved for 
\begin{equation}
m \approx -n \ln \left(1 - k^{-1}\right).
\end{equation}
This means, for example, that for large $n$ if we need to remove roughly half of the macroparticles we need $m \approx n\ln\left(2\right)$. This method is useful when the total charge/number of particles needs to be strictly conserved.

5. \textit{Energy-conservative thinning} (energyT) is a modification of the previously described \textit{number-conservative thinning}. According to this method we also select a random macroparticle  $m$ times, but do this with probability proportional to energy, i.e. the $i$-th macroparticle is selected with probability $e_i w_i /E$, where $e_i$ is the energy of the particle represented by the $i$-th macroparticle and $E = \sum e_i w_i$. The macroparticles that have been selected $c_i \neq 0$ times are assigned with the weight $c_i E/\left(e_i m\right)$ and the others are removed. One can check that this procedure complies with the principle of agnostic down-sampling and also strictly conserves the total energy $E$ in each cell. However, this method does not strictly conserve the total weight $W$. This method is useful when the total kinetic energy of particles needs to be strictly conserved.

6. \textit{Conservative thinning} (conserv). This method, proposed in \cite{gonoskov.arxiv.2020}, can be configured to preserve several invariants simultaneously. Each invariant ($A$) can be represented by a linear equation:
$A = \sum a_i w_i$, where $a_i$ amd $w_i$ are the contribution and the weight of the $i$-th particle, respectively. The conservation of several invariants sets a system of linear equations, where the number of variables (weights $w_i$) can be controlled by the number of particles involved in the thinning procedure. If the number of particles is greater than the number of invariants, the system is undetermined. It turns out that it is possible to find two solutions with one of the weights being equal to zero and others being positive, so that the probabilistic choice of one of these solutions results in an agnostic resampling that reduces the number of macroparticles by one. The procedure can be repeated several times for a given set of macroparticles to reduce the number of particles down to $n/k$ (assuming that it is still larger than the number of invariants). This method is useful when a number of physical properties of particles need to be strictly conserved. The method can be configured to preserve one or several conservation laws, as well as contributions to grid quantities (such as the charge and current density). In the latter case the procedure is governed by the location of nodes, which is of artificial nature. An interesting alternative is to preserve one or several first central moments of the particle distribution in coordinate and/or momentum space. One more aspect is that the method can be applied to either highly populated cells (independently, cell-by-cell) or dense clusters determined by some clustering method. Some examples and more details are given in Ref.~\cite{gonoskov.arxiv.2020}. In the version referred to as \textit{conserv} in the following consideration the method is applied to particles of the same kind within cells with more particles than quantities to be conserved, and is configured to preserve the total energy, all three components of momentum, the total charge/weight and the first central moments of particle distribution along all three coordinates (8 invariants in total). In one of the tests we use the \textit{conserv2} version that also preserves the second central moments of particle distribution in space (11 invariants in total).

For comparison we also consider methods that do not comply with the principle of agnostic down-sampling. Most of these methods revolve around merging of dense clusters.

7. \textit{Merging to averaged location} (mergeAv). According to this method, for each cell we determine $n_{cell}/k$ (where $n_{cell}$ is the number of particles in a cell), but at least 3, clusters using the k-means method with respect to location of particles in the momentum space. Their proximity in coordinate space is ensured by them being in the same cell. Next we replace all particles in each cluster with a new macroparticle that has the mean coordinate and momentum of particles in the cluster and weight equal to the total weight $W$. The number of selected clusters determines the number of macroparticles remaining after the merging procedure. Since the complexity of the k-means method is $({n_{cell}}^{3}/k)$, the algorithm may require significant computational resources.
This method is useful when the phase space can be adequately represented by a number of dense clusters. 
If $n_{cell}/k>>1$, the overall resulting number of macroparticles is approximately $n/k$. Due to computational time restrictions, the recommended value of $k$ for merge-based methods is such that no more than 30 clusters are formed.

8. \textit{Merging to random particle} (merge).
As we mentioned earlier, the merging procedure can naturally result in the systematic relocation of particles towards denser regions. An indicative example is the case of a bulk of particles (or a particle beam) with a narrow distribution in coordinate space (as compared to the coordinate scale of clusters used for merging). In this case merging naturally favors bringing macroparticles to the peak of that distribution, removing the macroparticles in its tails. We can avoid this by introducing the following modification to the previous method. The weight of particles in each determined cluster is brought to the location of a random particle in the cluster. Note, however, that this particular procedure does not prevent the reduction of particles' spread in the momentum space.

As a general note we would like to highlight the following. Any down-sampling results in the loss of information since the amount of unique macroparticles decreases. The inevitable consequence of this is the increase of noise in the distributions of particles. The more macroparticles are used the less noise we can expect and vise versa. In practice this leads to the trade-off between the accuracy of results and computational demands. The goal of arranging appropriate resampling is to avoid systematic deviations and minimize computational demands, and at the same time reach the accuracy necessary in the problem of interest.

To not spend computational resources for resampling when it is not needed, we use a trigger for starting the resampling procedure: the number of macroparticles in a shared-memory computational domain must reach a certain threshold value. All methods except \textit{globalLev} are applied to each cell independently.

\section{Comparison of resampling methods on test problems}

In this section we present the comparison of the resampling methods described above using two test problems: a steady-state plasma and the development of a Weibel instability in two counter-streaming plasma flows.

\subsection{Steady-state plasma}

In order to compare how strongly various resampling strategies can affect simulations we consider a 3D volume with a uniform steady-state plasma and compare the temperature change caused by a single instance of resampling which reduces the number of macroparticles by a given factor $k$. In our numerical experiments we observe a decrease of plasma temperature whether we use thinning or merging. To clarify the reasons of this observation, we start from a brief phenomenological analysis. 

For simplicity we consider the resampling of electrons, whereas the ions (or positrons) are modelled by a uniform positively charged background. We can outline two basic reasons for the change of temperature to happen due to resampling. Firstly, if we apply a non-agnostic resampling that does not preserve the total kinetic energy of particles, then we can potentially have an asymmetric net acquisition of energy mismatches, either positive or negative. This obviously happens in case of merging into a single particle because the particles selected for merging are likely to have thermal, isotropic distribution of momenta and their merging into a particle with mean momentum reduces the energy (fixing this with more advanced merging strategies is discussed in the end of the section). Secondly, even if the used resampling is agnostic, the change of weights effectively draws the plasma out of the equilibrium state. There could be two cases.

If the Poisson’s equation is solved at each iteration (as it happens in some spectral codes), an abrupt local relocation of charges builds up an additional local variation of electric field, the energy of which can eventually add up to a temperature increase. Note that if the resampling does not preserve charge within each cell, the local variation of charge density can build up a non-zero global electric field (especially in the 1D case), which can have a significant energy. This explains why preserving charge may be beneficial.

In the case where the Poisson’s equation is not solved and the field evolution is only driven by the charge currents, the local removal of macroparticles would effectively mean adding compensating charges that are fixed in space (the added positive charge is compensated by the increase of weight of the remaining macroparticles). In this case the plasma will tend to a new equilibrium state relative to the positively charged background with corresponding local variations of charge density. Again, if the charge is not preserved locally, a global effective potential variation may form and the placement of compensating charges may cause a significant energy change. Now, let us imagine a situation when the plasma (together with the initial positive background) leaves some part of the computational region. The added effective positive charges in this region will show up as unchanged noise in the electric field. Since this noise would contain a strictly positive energy, we can conclude that this energy is effectively deducted from the thermal motion of charges due to the application of resampling. Hence, resampling can cause an effective cooling of plasma. This is observed in our numerical experiments.

Particularly, we consider a 3D region represented by $32 \times 32 \times 32$ cells with periodic boundary conditions filled with homogeneous quasineutral electron-positron plasma with initial temperature $T_0=0.001mc^2$ (here $m$ is the electron charge, $c$ is the speed of light), cell size equal to 2 Debye radius, and \textit{physical} density derived from these values. The considered values of the initial number of macroparticles are $N_0=100$ and $N_0=1000$ particles per cell (ppc). The time step is set to $1/128$ of the period of cold plasma oscillations. After $t=1$ oscillation period we perform resampling, at $t=10$ oscillation periods we calculate the temperature of the plasma (as the average kinetic energy of particles) in comparison to the initial temperature $T_0$. This procedure was performed for values of $N_0$ mentioned above for every method of resampling and for a set of resampling coefficients $k$, which indicate the target decrease ratio in the amount of macroparticles, equal to (1.1; 3; 10; 30; 100; 300; 1000) where possible due to method limitations. To identify the temperature decrease induced by resampling the temperature decrease in the case without resampling is subtracted from the value obtained using various methods.

\begin{figure}[t!]
	\centering
	\includegraphics[width=0.49\columnwidth]{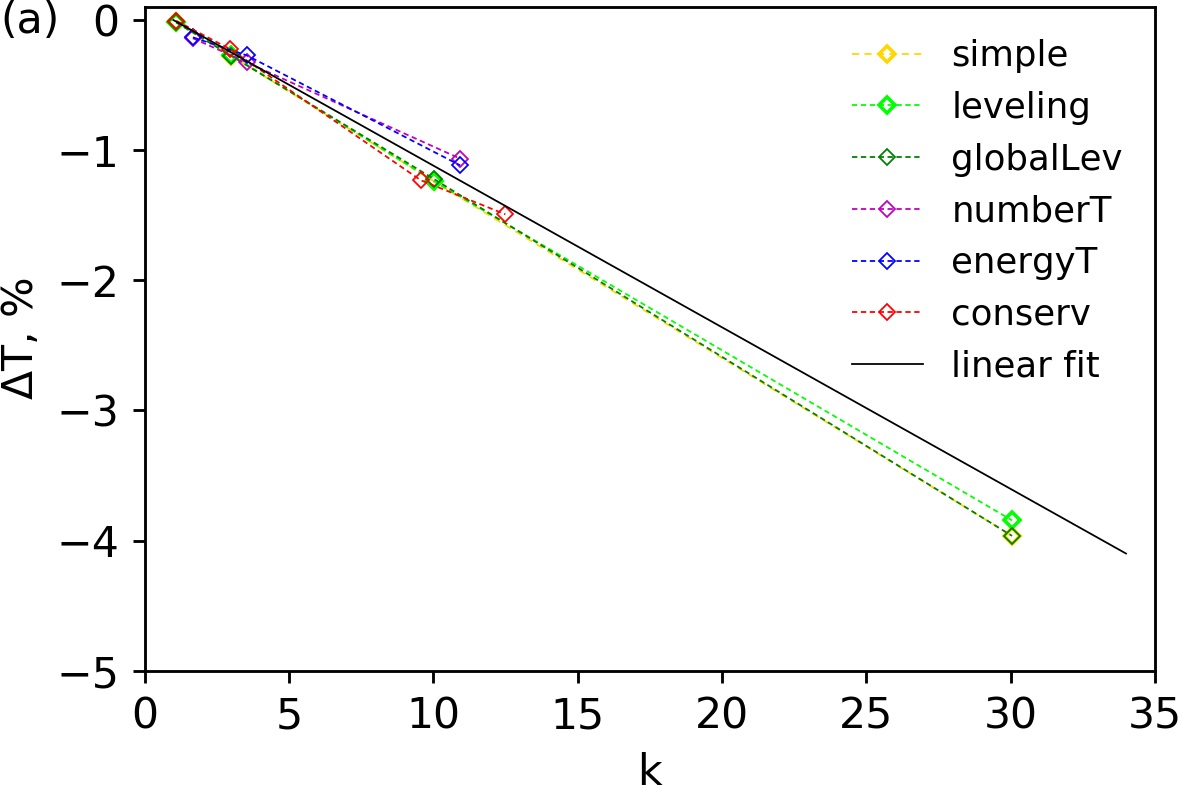}
	\includegraphics[width=0.49\columnwidth]{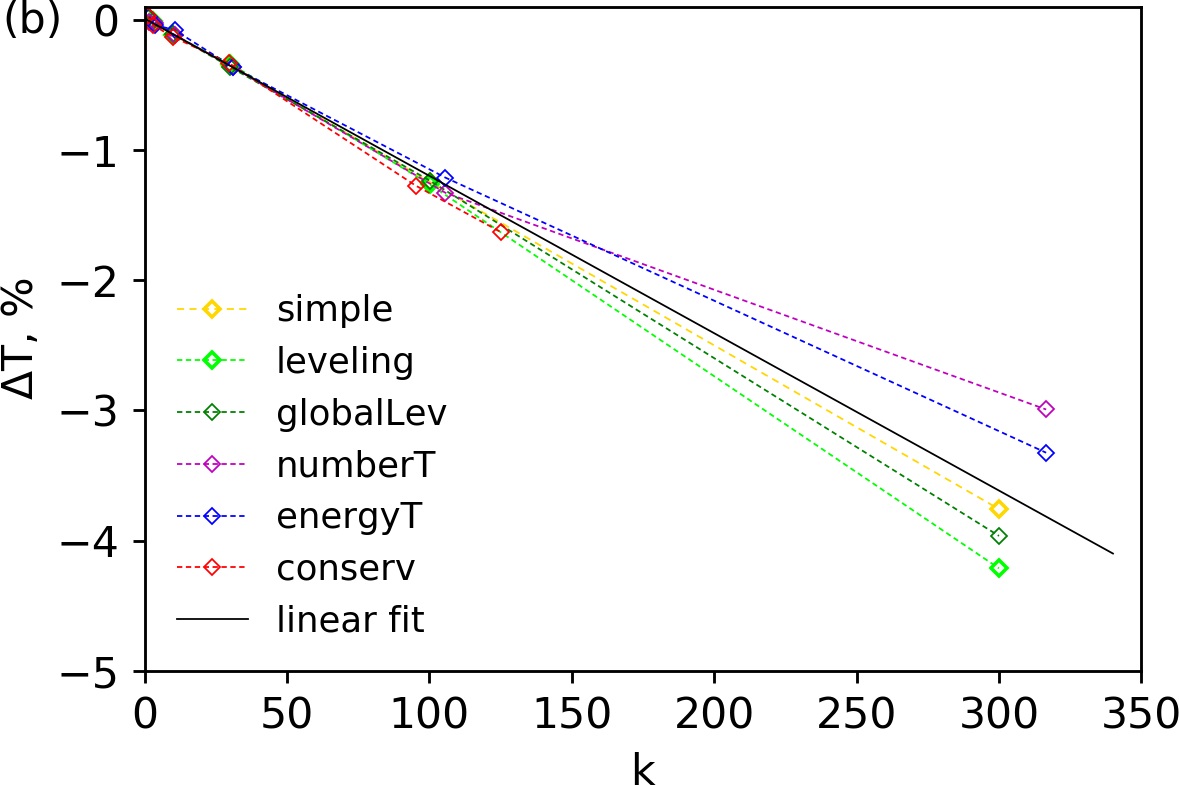}
	\caption{Change of equilibrium temperature for a plasma with initially (a) $N_0=100$ ppc, (b) $N_0=1000$ ppc after a single application of resampling depending on the resampling coefficient $k$. Only the results of thinning methods are presented. Dashed lines: methods' results, Solid: linear fit. Color online.}
	\label{dtemp1_2}
\end{figure}

\begin{figure}[t!]
	\centering
	\includegraphics[width=0.49\columnwidth]{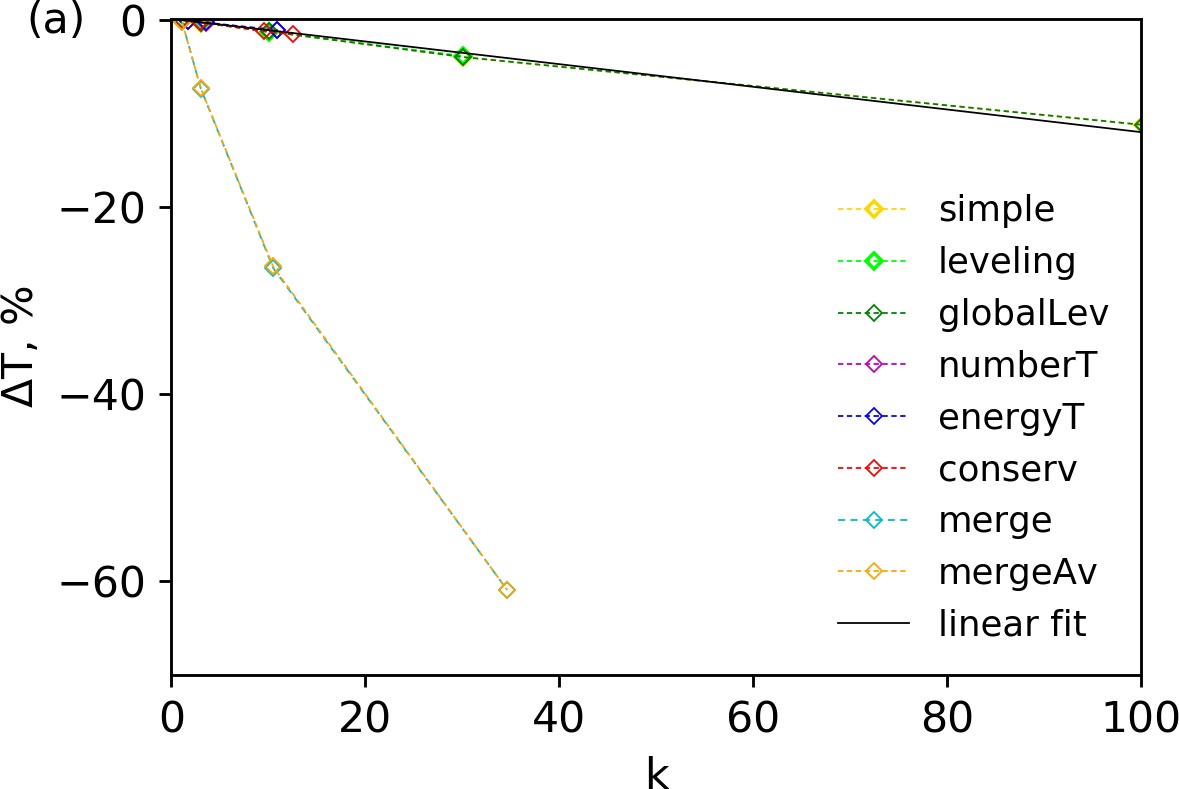}
	\includegraphics[width=0.49\columnwidth]{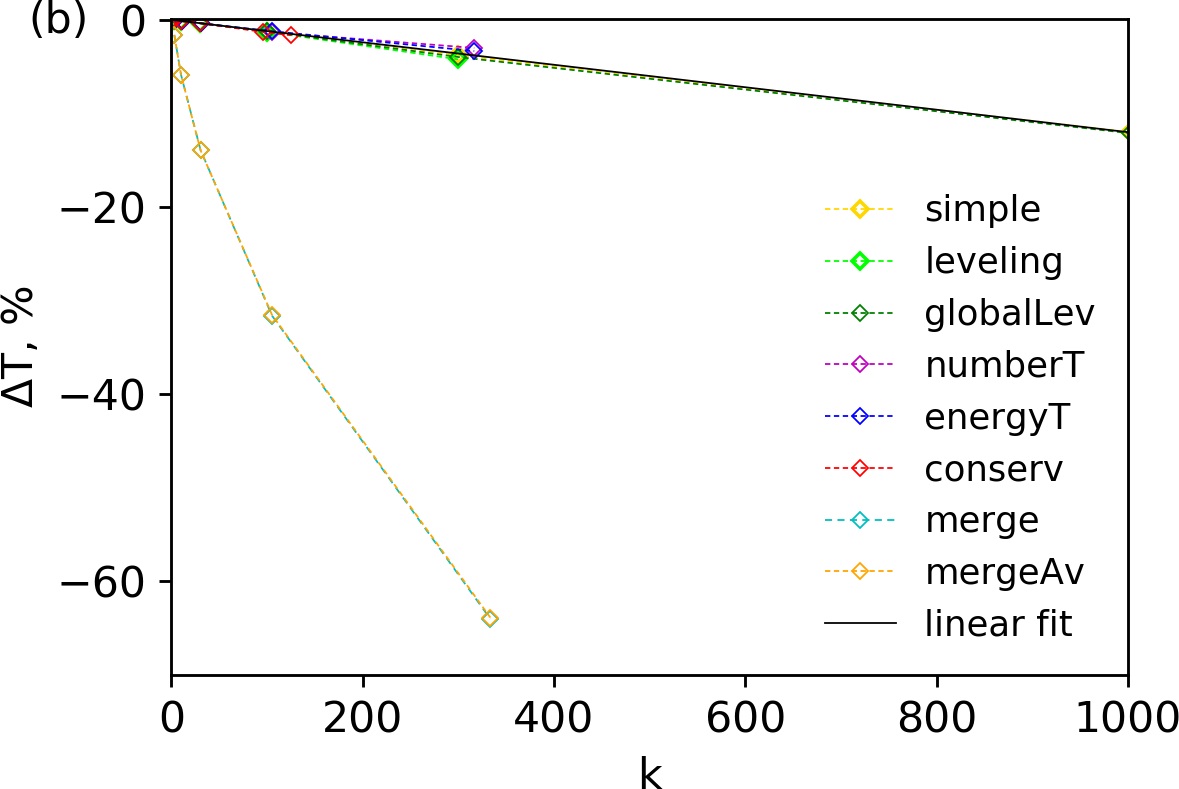}
	\caption{Change of equilibrium temperature for a plasma with initially (a) $N_0=100$ ppc, (b) $N_0=1000$ ppc after a single instance of resampling depending on the resampling coefficient $k$, \textbf{all} methods. Dashed lines: methods' results, Solid: linear fit. Color online.}
	\label{dtemp3_4}
\end{figure}

For all agnostic methods the results are similar and lie within a narrow range of each other, shown in Fig.~\ref{dtemp1_2}(a,b) for the cases $N_0=100$ and $N_0=1000$, respectively. Note that the values of $k$ in Fig.~\ref{dtemp1_2}-~\ref{dtemp3_4} represent the observed decrease ratio, which may deviate from the target due to methods' limitations.

Interestingly, the results show a linear trend in $\Delta T(k)$. In addition, close trends with respect to $N_0/k$ values (see Fig.~\ref{dtemp1_2}) indicate that the temperature decrease depends solely on the number of macroparticles per cell remaining after resampling rather than on the initial ppc number and the coefficient $k$ separately. The results of agnostic methods can be roughly fitted by $\Delta T=-(0.12/ppc_f)*T_0$, where $ppc_f$ is the final number of macroparticles per cell after resampling. Actual values of $\Delta T$ for each method vary by about 10\% from this rough estimate depending on the particular method in question and the value of k.

Merge-based methods, on the other hand, show considerably poorer performance according to our chosen metric (see Fig.~\ref{dtemp3_4}). Even in the best-case scenario where merge methods perform the closest to agnostic methods, $\Delta T$ shown by merge methods is approximately 15-20 times greater than $\Delta T$ shown by agnostic methods. Particularly, while agnostic methods show a linear trend towards a 10\% temperature drop at $k=N_0$ (which means the number of particles after resampling is $N_0/k\sim1$ ppc), merge methods yield a whole 65\% temperature decrease already at ppc~=~3. The curve for merge methods is slightly concave up, so for lower $k$ the result is even worse in relative comparison to the agnostic methods.

We see that in this particular case and according to the chosen metric, thinning outperforms merging. One can rightfully note that this is not surprising as we used an overly simple version of merging that repeatedly decreases the energy: if we consider one of the clusters of particles with a near-isotropic, thermal distribution of momenta, it is merged into a single particle having near-zero momentum. The use of merging into a pair of particles to conserve energy will obviously improve the results dramatically (this is a good reason to use the methods described in Refs.~\cite{vranic.cpc.2015, rjasanow.jcp.1996, rjasanow.jcp.1998}). Nevertheless, in this case the two particles appearing as a result of merging the selected group have the opposite momenta of the same absolute value that is determined by the total energy in the group. This may keep the temperature unchanged, but obviously deforms the Maxwellian distribution in such a way that the distribution of kinetic energy becomes peaked around the mean value. We could introduce another metric to quantify this and, probably, a more advanced version of merging could be designed with this metric in mind. The limit of this potentially long chain of improvements is to require the conservation of all the distribution functions, which is the kernel idea of the principle of agnostic resampling.

Let us now turn to the agnostic methods. Even the simplest versions that do not conserve total energy perform fairly well. This can be seen as an instructive demonstration of the fact that the agnostic principle itself enforces good performance even without configuring the method to conserve the quantity of interest.

Because of the presented arguments we will stick to the consideration of the simplest versions of merging in the following analysis. The reader should keep in mind that more advanced versions of merging can perform better in the considered cases. Although the merging algorithm can be tailored to specific conditions, we are interested in the situations when there is no sufficient understanding of the simulated process to perform such tailoring. We hope that the comparison of the versions of merging and thinning that are not tailored to the specific nature of problems can better indicate the principle differences between these approaches.

\subsection{Weibel instability in counter-streaming plasma flows}

One can argue that the previous test was favorable to thinning and unfavorable to merging as its non-conservative form was considered. Presumably, one of the main advantages of merging is that resampling is performed in a coordinated form that maintains the uniformity of the particle distribution. The next test is designed to examine this strong side of merging and identify the capabilities of thinning in this respect.

The second test problem is the development of Weibel instability \cite{Weibel} in counter-streaming plasma flows \cite{Fried}. This instability results in an exponential growth of perturbations in plasma density, current and magnetic field along the direction transverse to the plasma stream. To make our experiment robust we introduce a periodic modulation of density in the transverse direction to act as a systematic seed for the instability. For each method we carry out an individual simulation and perform a single resampling procedure near the beginning of the growth. In such a way we intend (1) to see whether the random perturbations caused by resampling can disturb the process, and (2) compare different resampling methods according to the extent of introduced perturbations. To quantify this extent of perturbations we measure the variance of plasma density computed for individual cells of the computational grid.

Let us first note that the reduction of the number of macroparticles should naturally result in the increase of variance for the number of physical particles in each cell. The extent of this increase, however, depends on the method. For example, \textit{number conservative thinning} does not immediately change the number of physical particles in any cell (although it changes after the migration of particles between the cells). In this case the impact of resampling on the ensemble of macroparticles is coordinated within each cell. To estimate the worst case scenario, let us consider the case of \textit{simple thinning}, for which the individual changes for macroparticles are totally uncoordinated. With probability $p_1=1/k$ the particle’s weight $w$ is increased by a factor of $k$ to $w_1=k w$, otherwise (with probability $p_2=1 - k^{-1}$) the macroparticle is deleted. The expected value of the number of physical particles $N_{phys}$ among different realizations of this random process must remain unchanged: $E[N_{phys}]=\frac{1}{k}w_1=w$ (hereafter by $E[\cdot]$ we denote the value averaged over all realizations of resampling). For the contribution of individual macroparticles, we can compute the variance $D[N_{phys}]=E[{N_{phys}}^2]-{E[N_{phys}]}^2$, and $E[{N_{phys}}^2]=\sum p_i w_i^2$, where $p_i$ is the probability of the $i$-th outcome and ${N_{phys}}_i = w_i$ is the number of physical particles in that outcome. For simple thinning we can write
$E[{N_{phys}}^2]= k^{-1} w_1^2 + (1 - k^{-1}) \cdot 0 = kw^2$. Finally, the variance is $D[N_{phys}]=kw^2-w^2=(k-1)w^2$. Since the variance is additive, considering a cell with $N$ macroparticles with weight $w$, we obtain: $D[N_{phys}]=N(k-1)w^2$. We are interested in macroparameters, such as physical particle density $n=N_{phys}\Delta V^{-1}=Nw \Delta V^{-1}$, where $\Delta V$ is the cell volume. We can calculate
\begin{equation}
D[n]=\frac{D[N_{phys}]}{(\Delta V)^2}=\frac{N(k-1)w^2}{(\Delta V)^2}=\frac{n(k-1)w}{\Delta V}=\frac{n^2(k-1)}{N}.
\label{eq:variance}
\end{equation}
Although this expression is the variance of density $n$ over different realizations of random events, the independence of such events in different cells allows us to use it to calculate the variance of $n$ over coordinate space.

For our study, we have performed several series of 2D simulations of the Weibel instability development in counter-streaming flows of quasineutral electron-ion plasma. We considered the following parameters: initial density $n_0=10^{22} $~cm$^{-3}$, plasma flow velocity $V_\pm=\pm 0.99995c$, which corresponds to a Lorentz-factor of $\gamma_0=100$, where $c$ is the light velocity and the "+" and "-" signs denote the streams directed along and opposite to the $x$ axis in our simulations.
The initial density of ions in both streams was set to be uniform. The initial density of electrons and their local momentum were modulated harmonically across the transverse direction ($y$ axis): $n_{\pm}=\pm \delta n_0 \cos(k_y y)$,
$p_{y,\pm}=\pm {\delta m_e {{\omega}_p}^2 {V_0}^2}\left(k_y\Gamma\right)^{-1} c^{-2} \sin(k_y y)$, where $\delta=0.02$ is the modulation amplitude, $k_y=2\pi/L$, $L=2.5\cdot10^{-5}$~cm is the spatial scale of the modulation, ${\omega}_p=\sqrt{8\pi e^2n_0/m_e}$ is the plasma frequency (of total density of both streams), $m_e$ and $e$ are the electron mass and charge. In the considered case of small-scale spatial modulation, the growth rate of the Weibel instability is $\Gamma \approx {\omega}_p \gamma_0^{-1/2} {V_0}/{c}$. The modulation of electron density leads to the electromagnetic field variation of the following form:
$\vec{B}=8\pi\delta\frac{en_0}{k_y}\frac{V_{+}}{c}\sin(k_y y)\vec{z_0}$,
$\vec{E}=-8\pi\delta\frac{\Gamma}{ck_y}\frac{en_0}{k_y}\frac{V_{+}}{c}\cos(k_y y)\vec{y_0}$.

In our simulations the size of the simulation region is set to $2 \: \mu $m$ \times 4 \: \mu$m ($96\times384$ cells) with periodic boundary conditions and the time step equal to ${1}/(128\Gamma)$. The computation time is set to match the duration of the linear regime during which the plasma density perturbation is negligible compared to the plasma density itself, which was the case until $t \approx 3.75/\Gamma$. The computation is performed for each method of resampling and each value of the resampling coefficient $k$ from the set $(1.1;2;5;10;20;50)$, as well as for the case without resampling. (The methods \textit{merge} and \textit{mergeAv} could not complete for coefficients $k=1.1$ and $k=2$ due to computational time restrictions.) In every computation the resampling procedure is applied once at $t=1.25/\Gamma$. In order to identify the variance induced by the resampling procedure, for each method and each value of $k$ the difference $\Delta D(t) = D[n](t) - D_0[n](t)$ is calculated, where $D[n](t)$ is the time dependence of the variance of particle density $n$ in that calculation and $D_0[n](t) \sim e^{2\Gamma t}$ is the variance of physical density in the simulation performed without resampling.

\begin{figure}[t!]
	\centering
	\includegraphics[width=0.49\columnwidth]{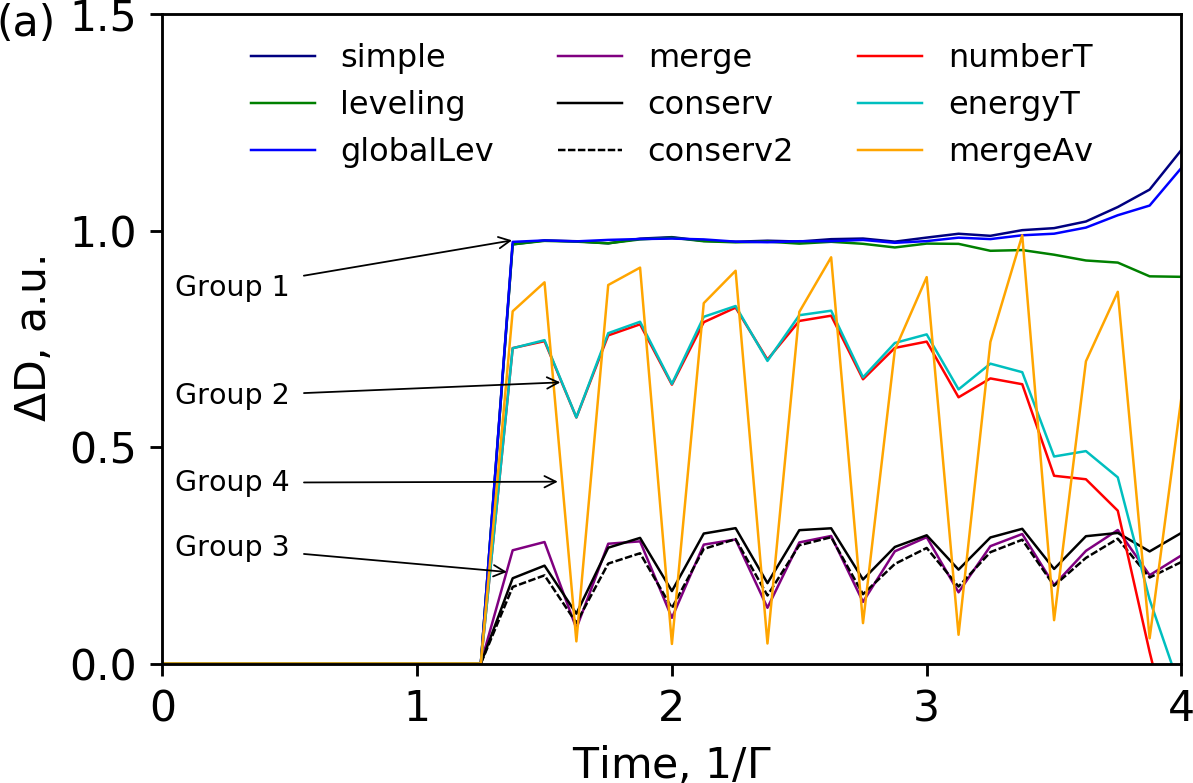}
	\includegraphics[width=0.49\columnwidth]{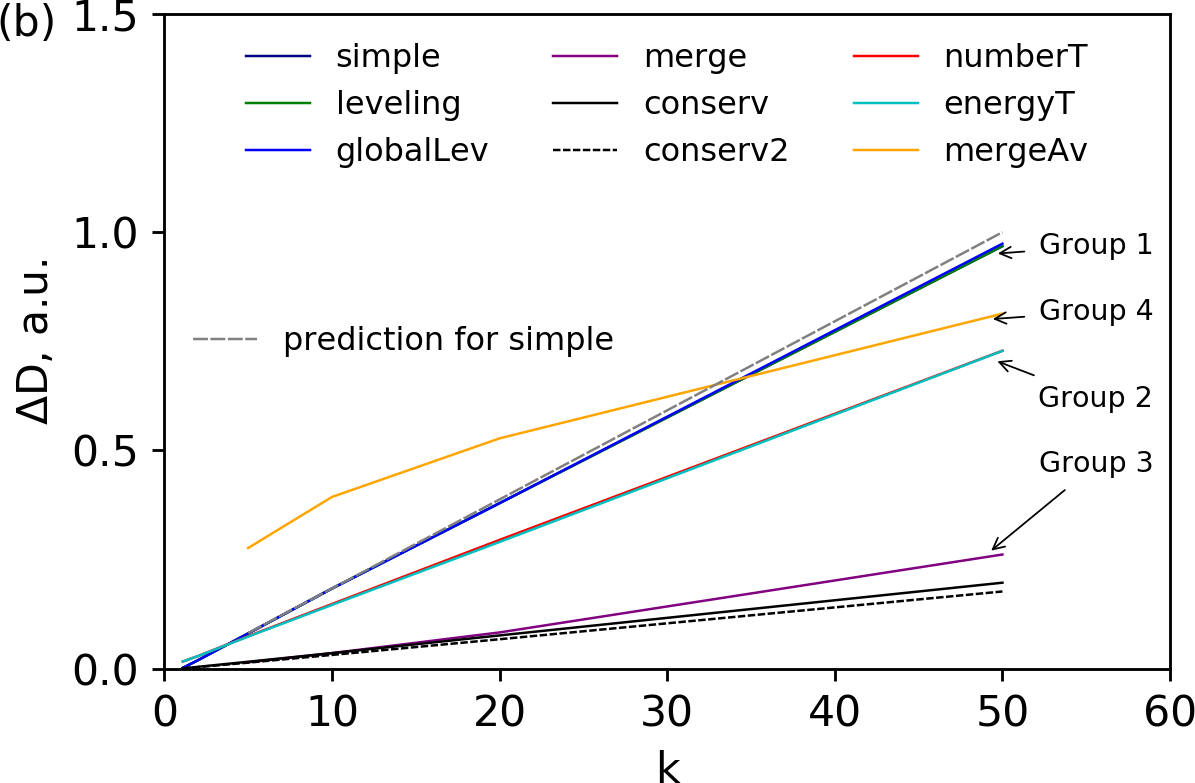}
	\caption{Variance of physical particle density induced by different methods of resampling $\Delta D$ normalized to theoretical result (Eq.~\ref{eq:variance}). (a) Temporal evolution $\Delta D(t)$ for target resampling coefficient $k=50$. (b) Dependence of D on target resampling coefficient $\Delta D(k)$ immediately after resampling. All methods can be grouped into four groups.
	Group 1: \textit{simple} (dark blue), \textit{leveling} (green), \textit{globalLev} (blue); Group 2: \textit{numberT} (red), \textit{energyT} (cyan); Group 3: \textit{merge} (purple), \textit{conserv} (black), \textit{conserv2} (dashed black); Group 4: \textit{mergeAv} (orange). Color online.}
	\label{dD_50}
\end{figure}

Let us first compare the results of different methods using equal values of $k=50$, see Fig.~\ref{dD_50}(a). According to these results, the methods can be divided into 3 groups. Within each group the results are very similar. Specifically, for methods in Group 1 (\textit{simple, leveling, globalLev}) $\Delta D(t)$ takes the form close to a step function. In other words, resampling causes a single leap in $\Delta D(t)$ at the time of resampling, after that the difference in variance of physical density compared to the case without resampling stays constant up until the end of the linear regime, despite the fact that the variance itself grows exponentially. This result confirms that in this case the influence of resampling on the variance of physical density can be considered independently of other (physical) processes affecting it.

For methods in Group 2 (\textit{numberT, energyT}) the results are of the same order of magnitude and follow a similar trend, but these results show a notable periodic oscillation in $\Delta D(t)$, which is not negligible, but still of low amplitude compared to the value itself.

Group 3 methods (\textit{merge, conserv, conserv2}) show a considerably lower $\Delta D(t)$. The results of the \textit{mergeAv} method are in between and thus this method forms a separate Group 4. For methods in Groups 3 and 4 the oscillations are of the same order of magnitude as the value itself.

Now let us compare the results of each method depending on the value of $k$. In order to assess that, we present values of $\Delta D(t)$ immediately after resampling. Since before resampling all computations in our series are identical, this value is exactly the leap in variance caused by resampling. In Fig.~\ref{dD_50}(b) we present this value as a function of the target resampling coefficient $k$ for all resampling methods, as well as the estimate (\ref{eq:variance}) for the simple thinning method.

As evident, Group 1 results follow the estimate (\ref{eq:variance}): a linear dependence proportional to $k-1$. The results in Group 2 have a slightly lower slope: compared to Group 1 the variance increase is overall lower, with except of some range of low values of $k$. 
Group 3 methods performed close to linear, and yield notably lower increase than all other methods: the slope is $\sim 0.2-0.25$ of the slope for Group 1 methods. The \textit{mergeAv} method stands out as highly nonlinear, yielding higher increase of variance at low coefficients, but for $k=50$ the increase is lower than that of Group 1.

As anticipated, the \textit{merge} method performs better than the majority of thinning methods in this test. This can be attributed to the fact that this method provides the most advanced, coordinated replacement of macroparticles giving, in some way, the most efficient representation of the ensemble. Nevertheless, we should highlight that this is possible because we know the coordinate and momentum scales of the simulated process and ensure that they are larger than that of the merge procedure. A notable result is that the \textit{conserv} method performs equally well, even though it does not require any prior knowledge of this type. In the \textit{conserv} method the preservation of spatial uniformity is enhanced by preserving the first central moment (along each coordinate) of the particle distribution in the selected group. To verify that this is the cause, we perform an additional simulation employing the \textit{conserv2} method that also preserves the second central moments of particle distribution in space. The observed further improvement (although minor) supports our interpretation. This observation indicates that within the approach of agnostic conservative resampling one can maintain the uniformity by preserving one or several central moments of particle distribution. Note that a larger number of preserved quantities requires more particles in the group for resampling to be possible, whereas a smaller number of preserved quantities makes it possible to apply the algorithm to smaller groups being more localized in the phase space. This leads to an interesting dilemma of what the optimal strategy is.

\section{Comparison of methods on pertinent physical problems}

A large class of tasks that require resampling methods is the study of cascaded production of electron-positron pairs and high-energy photons.
in laser fields of high intensities. The transitions between the quantum states can be characterized by the rates computed within quantum electrodynamics (QED) and thus these cascades are commonly referred to as QED cascades. One of the most widely used numerical approaches for the simulation of QED cascades is based on extended PIC codes, also known as QED-PIC codes \cite{nerush.prl.2011, elkina.prstab.2011, sokolov.pop.2011, ridgers.jcp.2014, gonoskov.pre.2015, chang.2017}. With the development of an electrodynamic cascade, the number of particles can increase exponentially in time and significantly increase computational demands. This means that resampling will have to be applied on a continuous basis, in contrast to the case of test problems, where resampling was applied only once in each computation. It also means that some problems are impossible to compute without the use of resampling, so there may be no benchmark to compare results to. In this case we have to rely entirely on results acquired using one or more methods of resampling.

A cascade can have two stages: linear and nonlinear. At the linear stage, the density of the generated electron-positron plasma is not sufficient to significantly affect the structure of the electromagnetic field and its intensity. At the nonlinear stage, on the contrary, the generated plasma has a higher density and significantly affects the electromagnetic field. Therefore, at the linear stage resampling can affect only the particle distribution function, also potentially affecting the local rate of cascade development. At the nonlinear stage resampling may also affect the structure and magnitude of the fields, changing the plasma-field dynamics. Below, using several examples, we consider various stages of the QED cascade and show how different types of resampling can affect the simulated processes. For simulations, we used the PICADOR code with the \textit{Adaptive Event Generator} module described in \cite{gonoskov.pre.2015}, which, when necessary, subdivides the time step in order to resolve the QED cascade. 

\subsection{Linear cascade in a standing linearly-polarized plane wave}

To investigate the operation of resampling methods at the linear stage of the cascade, we chose a well-studied problem of QED-cascade development in the field of a  standing linearly polarized plane wave (see, for example, \cite {kirk_2009, BashmakovPP2014, GrismayerPRE2017s}). In order to have a benchmark for comparison (the results of simulation without the use of resampling, labeled with 'w/o', are possible in this particular case) we chose a relatively small wave amplitude $E_0=1000 m\omega_0c/e$, where $\omega_0=2.35\times10^{15}$~s$^{-1}$ is the laser frequency for wavelength $\lambda_0=0.8$~$\mu$m. The only non-zero field components are $E_z$ and $B_y$; this causes particles to move along the $x$ and $z$ axes. At this wave amplitude, electrons and positrons tend to the vicinity of the nodes of the electric field, but due to the stochasticity of photon emission and decay into pairs, electrons and positrons can reach the vicinity of the electric field antinode \cite{BashinovQE2018}.

Initially, electrons and positrons with the density of approximately $10^{17}$cm$^{-3}$ are uniformly distributed in a $\lambda_0\times\lambda_0\times\lambda_0$ simulation box represented by $128\times2\times2$ cells. The initial number of particles of each type in a cell was approximately 976. Initially, there are no photons, but they can be emitted by electrons and positrons. The boundary conditions for the fields and particles are periodic. The development of the cascade is considered during a $7T$ time interval, where $T=2.66$~fs is the wave period. The time step was $dt=1.33\times10^{-17}$s.

The same parameters are used for all resampling methods: every second iteration if the amount of macroparticles of any type exceeds the resampling threshold of $5\times 10^5$ particles, the macroparticles of that type undergo resampling with $k=2$, and the amount of these macroparticles decreases by approximately half. 
The simulations are carried out on the Broadwell section of the MVS-10P supercomputer of the Joint Supercomputer Center of RAS. In all runs we employ 1 computational node (32 cores overall) covering 1 computational domain.

\begin{figure}[t!]
	\centering
	\includegraphics[width=0.3\columnwidth]{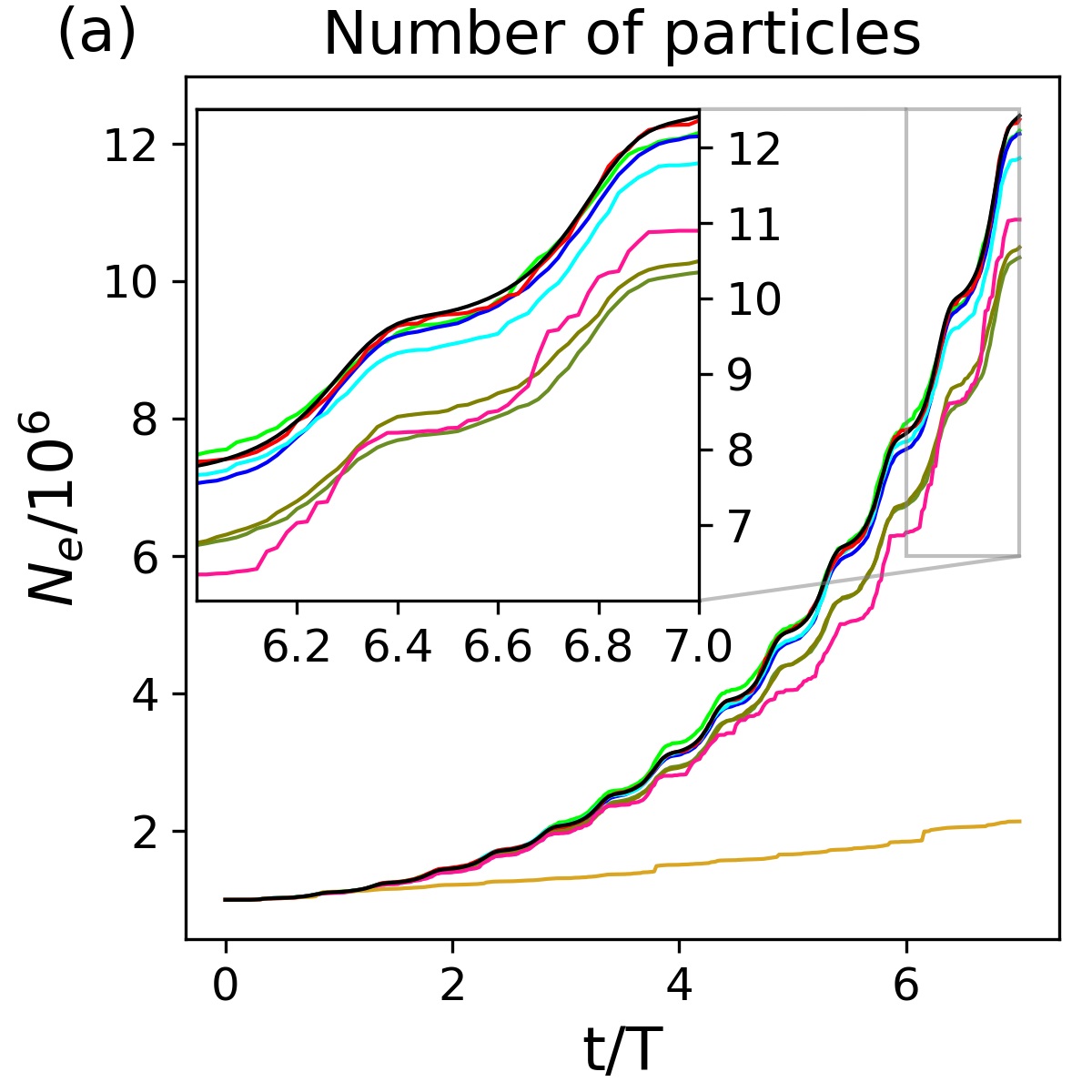}
	\includegraphics[width=0.3\columnwidth]{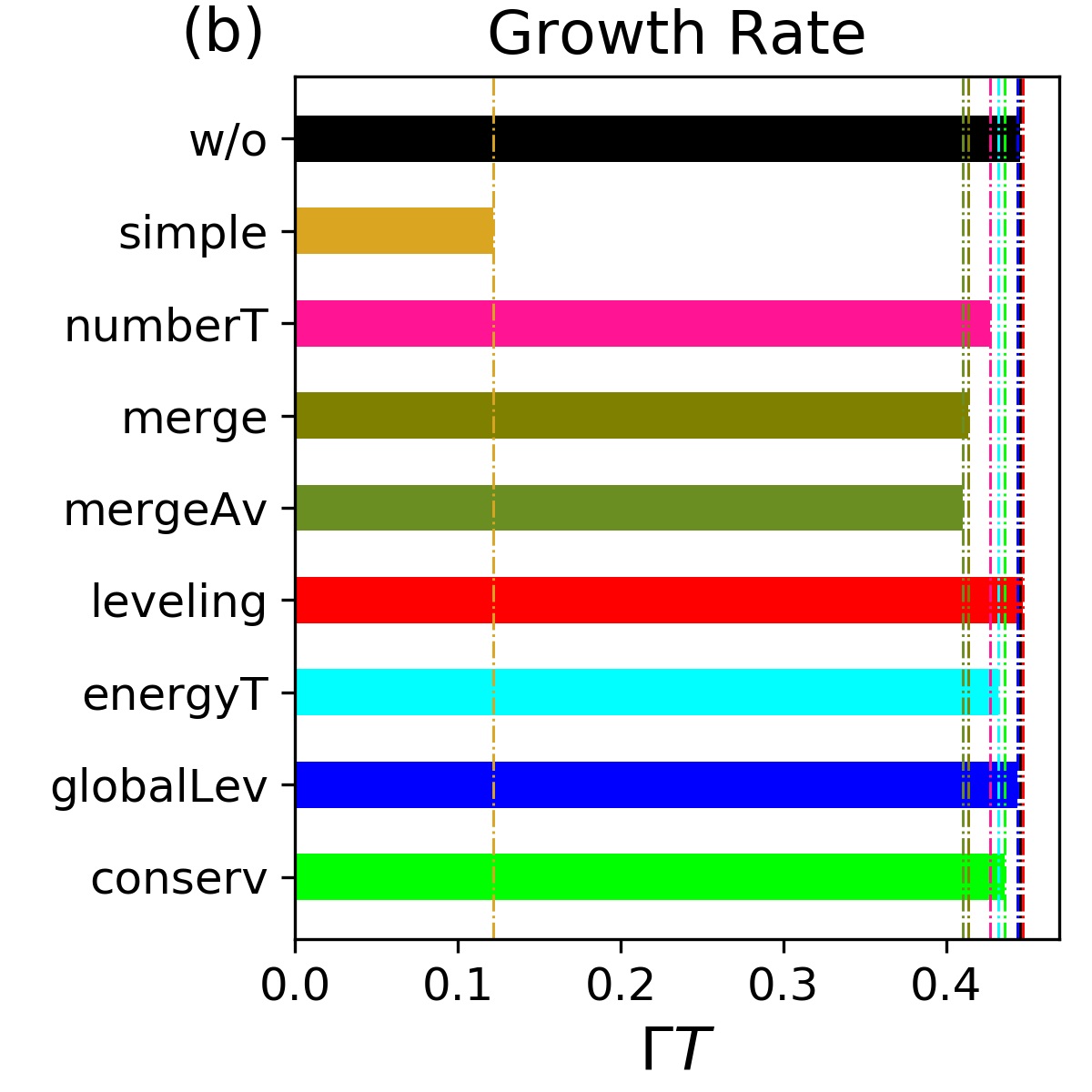}
	\includegraphics[width=0.3\columnwidth]{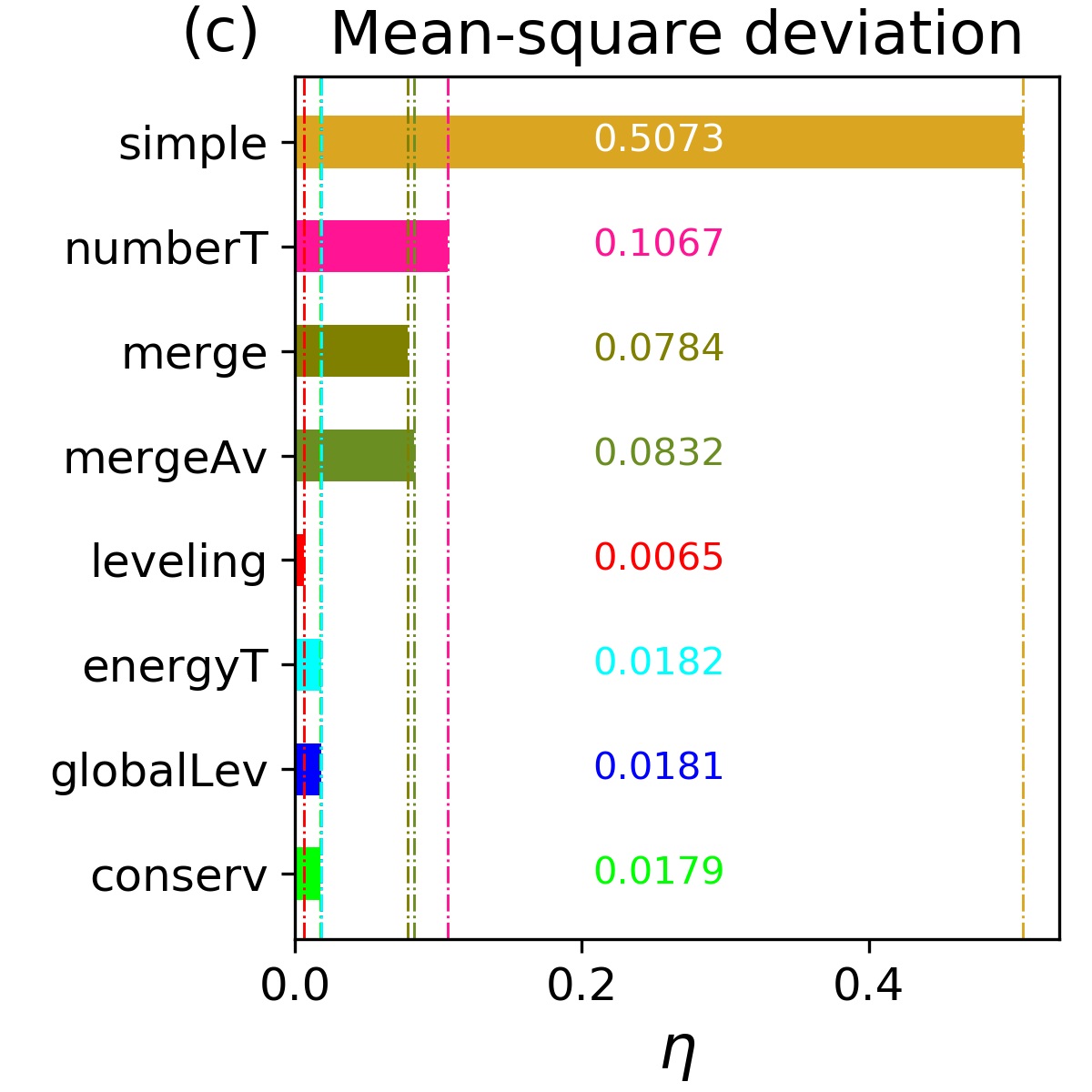}
	\caption{Comparison of different resampling methods via simulations of QED cascade development in a standing linearly-polarized plane wave. (a) Time dependence of the total number of electrons and positrons $ N_e $ in the simulation box. (b) The cascade growth rate $ \Gamma $. (c) The relative mean square deviation $\eta$ of $N_e(t)$ acquired with use of resampling from $N_e(t)$ acquired without resampling. The colors in figures (b) and (c) correspond to the color of the lines in figure (a). Color online.}
	\label{FCsGr}
\end{figure}

To analyze the effect of resampling on the accuracy of each simulation, we consider the temporal evolution of the total number of electrons and positrons $N_e$ (Fig.~\ref{FCsGr}(a)). For our analysis we use the following parameters: the cascade growth rate $\Gamma= \left(\ln\left(N_e\left(t = 7T\right)\right) - \ln\left(N_e\left(t = 3T\right)\right)\right)/\left(4T\right)$ and the relative mean-square deviation of $N_e(t)$ acquired with resampling from $N_e(t)_{w/o}$:
\begin{equation}
\eta =\sqrt{\frac{\int_0^{7T}\frac{(N_{e,res}-N_{e,w/o})^2}{N_{e,w/o}^2}dt}{7T}}, 
\end{equation}
where $res$ denotes the used type of resampling. In order to calculate $\Gamma$, the moment $t=3T$ is used as the initial one, since at $t=0$ there are no photons and it takes about $2.5T$ for the steady exponential growth to be established. 

Note that the principle of agnostic resampling ensures that the amount of physical particles is not affected on average. However, to obtain the growth rate one needs to calculate the logarithm of the number of particles, which is subject to random deviations $N + \Delta N$. This means that one should calculate the \textit{geometric} mean value of the growth rates obtained within several identical simulations (with different random number generator seeds), because the calculation of the \textit{arithmetic} mean value would yield a systematically lower value (e.g. $\left(\ln(N + \Delta N) + \ln(N -\Delta N)\right)/2 = \ln\left(\left(N^2 - \Delta N^2\right)^{1/2}\right) < \ln(N)$). Therefore, we see that for an appropriate calculation we need to know that the process is exponential. We intentionally use the arithmetic mean density value (over two runs) to mimic a situation when we have no prior knowledge about the process. In such a way we quantify the resampling-related error in the computation of outlined parameters, assuming that we do these calculations without any prior knowledge about the process. This also quantifies possible resampling-related distortions in more complex (non-linear) processes as we explicitly observe in other examples (see section 5.3.).

It should be noted that at a given wave amplitude, electrons and positrons emit plenty of photons, but only a small fraction of all photons decays into electron-positron pairs. Therefore, during the whole simulation resampling was initiated about 100 times for photons, but only once for electrons and positrons (Table~\ref{TCsGr}). Thus, hefty macroparticles are added to the ensemble of electrons and positrons as the result of photon decay. The simulation without resampling shows that $\Gamma T=0.445$ and the number of electrons and positrons increased by about a factor of 12 during the whole simulation (Fig ~\ref{FCsGr}(a)).

Based on the comparison of Fig.~\ref{FCsGr}~(a), (b) and (c), all considered resampling methods can be divided into 4 groups. The first group consists of the most accurate thinning method \textit{leveling} with $\eta=0.006$ and an error in the value of $\Gamma$ of about 0.3\%. The second group also includes fairly accurate thinning methods \textit{globalLev}, \textit{conserv} and \textit{energyT}, for which $\eta\approx0.018$, and $\Gamma$ can be determined with an accuracy of 0.3\%, 2.2\% and 3\%, respectively. The third group includes merge methods \textit{merge}, \textit{mergeAv} and the thinning method \textit{numberT} with $\eta\approx0.08; 0.08; 0.1$ and an accuracy of $\Gamma$ equal to 7.2\%, 7.9\% and 4\%, respectively. The fourth group consists of the most inaccurate thinning method 'simple'. This method within the performed simulations yields $\eta=0.5$ and an error in estimation of $\Gamma$ of 72\%, thus the accuracy of this method is unacceptably low.

Apart from the accuracy of simulations of physical processes, a computational speedup provided by different resampling methods must also be considered (Table~\ref{TCsGr}). Without the use of resampling the simulation run time was approximately 22000 seconds. The run time using thinning methods was approximately 1000 seconds, most of which is owing to the resampling of photons. In general, thinning methods speed up the simulation by about a factor of 20. At the same time, the run time using the \textit{merge} and \textit{mergeAv} methods is significantly higher, up to about 11000 seconds. This is primarily due to the use of the k-means method for merging of particles, and also due to a more frequent triggering of resampling. The merge methods yield a speedup of about 2 times in comparison to the 'w/o' simulation, and they have the same or even worse accuracy than thinning methods (with the exception of \textit{simple} thinning).

\begin{table}[t!]
\begin{center}
\caption{The influence of different types of resampling on run time and frequency of resampling.}
\label{TCsGr}
\begin{tabular}{ c | c | c | c  }
\hline
\textbf{\thead{Type of \\ resampling}} & \textbf{Run Time, s} & \textbf{\thead{Resampling of \\ Photons}} & \textbf{\thead{Resampling of \\$e^-$,$e^+$}} \\ 
\hline
\textit{conserv} & 1101.7 & 215 & 1 \\
\hline
\textit{globalLev} & 1060.5 & 207 & 1 \\
\hline
\textit{energyT} & 915.4 & 99 & 1 \\
\hline
\textit{leveling} & 965.7 & 134 & 1 \\
\hline
\textit{mergeAv} & 11274.5 & 264 & 1 \\
\hline
\textit{merge} & 11367.1 & 265 & 1 \\
\hline
\textit{numberT} & 848.3 & 96 & 1 \\
\hline
\textit{simple} & 978.9 & 214 & 1 \\
\hline
w/o & 21929.6 & 0 & 0 \\
\hline

\end{tabular}

\end{center}

\end{table}

Note that the large computational costs of the \textit{merge} method should be attributed to the use of the $k$-means algorithm, whereas other, less demanding approaches could also be applied. In this context one should differentiate two essential components: the algorithm of merging a given group of particles (this concerns the conservation laws) and the algorithm of selecting the particles to be merged. The latter is commonly referred to as clustering and can be implemented in many different ways (see, e.g. \cite{chang.2017, luu.cpc.2015, vranic.cpc.2015, rjasanow.jcp.1998, martin.jcp.2016}) so that the computational costs can be much lower than that of the $k$-means algorithm. For example, in Refs.~\cite{chang.2017, vranic.cpc.2015} the authors distribute particles of each cell in coordinate space among the cells of an additionally introduced grid in momentum space and select the cells that contain three or more particles. Note, however, that this requires a prior knowledge about the scale of acceptable difference in momentum of particles to be merged. This imposes a restriction: with reduction of this acceptable difference, the chance of finding cells with three or more particles decreases and this effectively means that the algorithm starts to require more particles per cell (in coordinate space) to function, which may limit the capabilities of the algorithm to perform down-sampling. The use of the $k$-means algorithm also implies the presence of the acceptable scale for the difference in momentum space, but this scale is adaptively selected (independently for each cell) by the logic of this algorithm. This makes this approach more flexible but less controllable. Finally, we should note that it is also possible to apply thinning to the groups selected by some clustering algorithm. In this case, if the thinning algorithm is agnostic, it is no longer absolutely necessary to control the coordinate-momentum scales of clustering, but the localization of the clusters in the phase space can still contribute to the quality of resampling \cite{rjasanow.jcp.1996, rjasanow.jcp.1998}.

\subsection{Nonlinear cascade in a linearly-polarized standing wave}
In this problem we perform a full 3D3P simulation of irradiation of a seed plasma target by two colliding half-infinite linearly-polarized ultraintense laser pulses. As shown previously \cite{Muraviev.jetplet.2015}, in such a configuration the electromagnetic cascade \cite{nerush.prl.2011, elkina.prstab.2011, sokolov.pop.2011, ridgers.jcp.2014, gonoskov.pre.2015, chang.2017} can result in a rapid generation of plasma during the cascade's linear stage followed by the formation of ultra-thin current sheets during the nonlinear stage. We perform the simulation of this problem using different methods of resampling. Performing it without resampling is not possible due to the inherent strain on computational resources.

It is important to note the following. In this setup, since the plasma density can grow significantly over a single half-period of the electromagnetic wave, the setup is highly sensitive to the phase of the standing electromagnetic wave at the instant when the plasma density becomes sufficient to affect the field structure of the standing wave. Consequently, the probabilistic nature of electromagnetic cascading may lead to a discrepancy in results (in physical and numerical experiments alike) over different realizations of probabilistic \textit{physical} events. We stress that a slight discrepancy in observed parameters in calculations with the same initial conditions may represent different realizations of a probabilistic \textit{physical} process and it alone does not indicate that the resampling process leads to the distortion of the result.

The chosen initial parameters for this problem are: the laser wavelength $\lambda = 800$~nm, the field amplitude is $a_0=3500$ in relativistic units, the laser beam width $D=5\lambda$, the initial seed plasma represents a $1.5\lambda\times0.5\lambda\times0.2\lambda$ block with density equal to $10^{-3}$ of the critical plasma density $N_{cr}$. The simulations are carried out on the Petastream section of the MVS-10P supercomputer of the Joint Supercomputer Center of RAS. In all runs we employ 8 computational nodes (64 cores overall), 1 MPI process per node, 2 OpenMP threads per core. The 3D simulation area is decomposed into 2 computational domains in each direction.

\begin{figure}
	\centering
	\includegraphics[width=0.7\columnwidth]{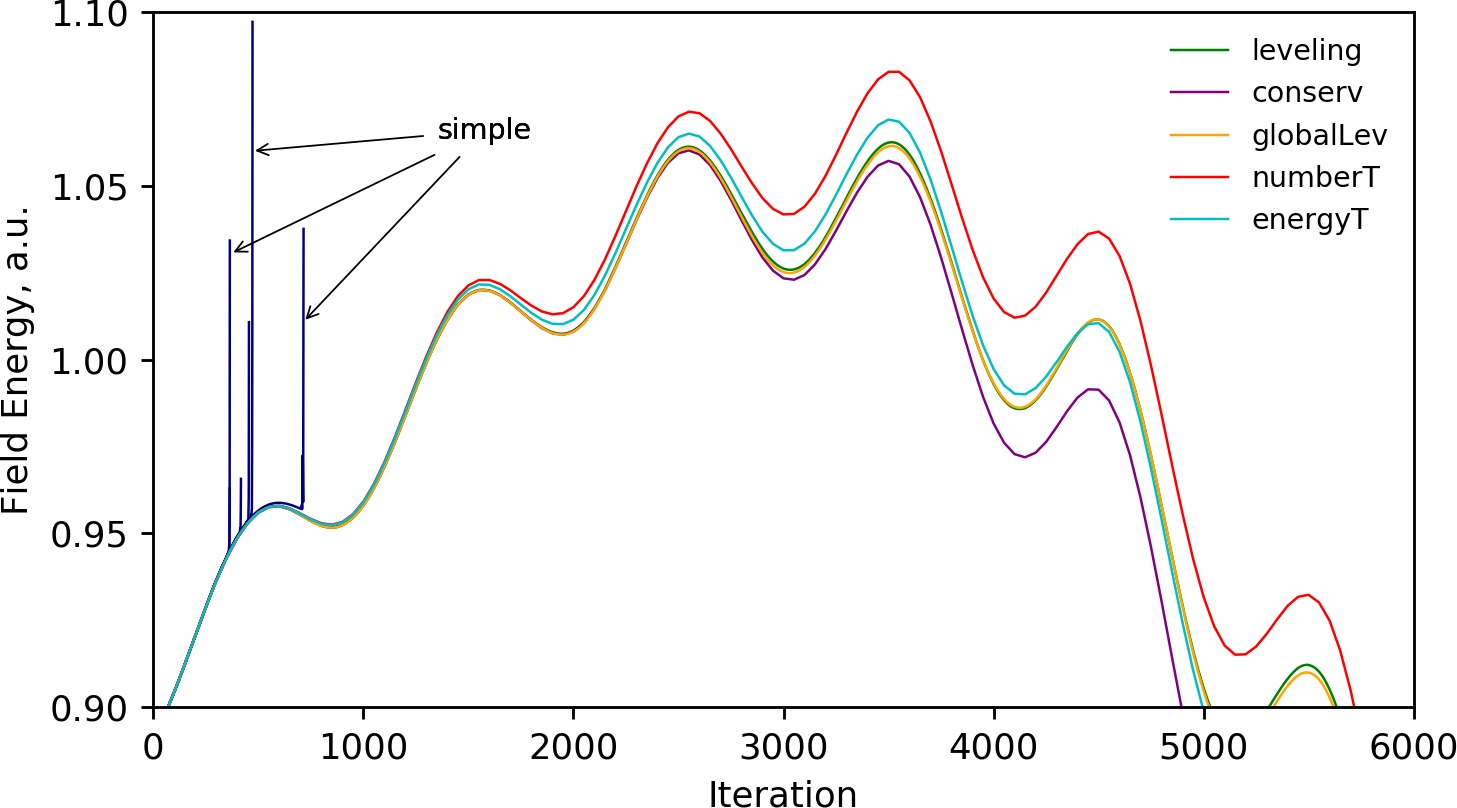}
	\caption{Total Field Energy as a function of time in simulations of a QED cascade in a linearly-polarized standing wave computed using different methods of resampling. Color online.}
	\label{SheetsFE}
\end{figure}

In Fig.~\ref{SheetsFE} we present the total electromagnetic field energy in the $7\lambda\times8\lambda\times8\lambda$ simulation box for different methods of resampling as a function of time since the start of the nonlinear stage, when the current sheets begin to form. Thinning methods were compared using the same default value $k=2$, which we believe to be optimal. This value of $k$ is well outside the recommended range for merge-based methods dictated by computational time restrictions, so here these methods were not considered. 

As shown in Fig.~\ref{SheetsFE}, agnostic methods excluding \textit{simple} yield qualitatively similar results varying within $5\%$, thus cross-confirming each other's results. For the method \textit{simple} we performed five attempts with different seeds of the random number generator. Nevertheless, all these attempts have been terminated at an early stage with a clearly unphysical surge in the field energy. 

Deeper investigation of the performance of the method \textit{simple} shows that although \textit{average} density is conserved well (up to a point), \textit{maximal} density is not (see Sec. 5.3 for more detail). Since each particle's weight is probabilistically increased or zeroed independently, \textit{simple} resampling with coefficient $k$ may result in the total weight in a certain cell increasing by a factor of up to $k$. This phenomena is a lot more likely to occur if much of that cell's total weight is carried by a single particle or few particles. If $k$ is large enough, or if the procedure is applied many times (our case), this can result in a significant increase of the total particle weight in some cells (effectively at the cost of other cells).
Since on average properties are conserved, intuition may suggest this should not lead to physically incorrect results, apart from some increased numerical noise. However, it can be shown that this is not the case. The reason behind this lies in the discrete nature of the particle-in-cell code. In order for it to operate correctly, all physically relevant time scales have to be resolved by the PIC time step. The artificial increase of total weight (and thus physical density) in some cells leads to the increase of the local plasma oscillation frequency. If this frequency exceeds the frequency resolved by the PIC code's time step, an unphysical instability develops. Particularly, the current generated by a very hefty particle or cluster of particles can be so strong that it induces an electric field that inverts the momentum of these particles on the next iteration. The process is additionally fed by new particles created in an electromagnetic cascade, which results in an unphysical exponential growth in particle density, energy and field values, resulting in a termination of the computation, often in as few as several iterations.
Other methods either strictly conserve certain values or exclude particles with a large weight from the procedure, helping to avoid this problem.

To summarize, we see that the \textit{simple} thinning method may not be applicable if a local stochastic variation of density can cause numerical instability. All other thinning methods show adequate results similar to each other in this setup.

\subsection{Pinching of electron-positron plasma in a multi-10PW dipole wave}

To study the effect of various resampling techniques on the dynamics of the
particle ensemble in another pertinent problem, we have investigated the 
interaction of a
multi-10PW-level laser radiation with plasma targets. Such a problem is of great interest due to the fact that this kind of laser system will soon be available to be used in experiments ~\cite{ELI,APOLLON,VULCAN}
and even more powerful 
systems are being developed~\cite{SEL,XCELS}. 
These experimental setups will 
allow the studies of QED cascade development in converging fields of petawatt power level. Moreover, these laser systems will be capable of driving various self-consistent nonlinear regimes 
during the interaction of a QED-produced electron-positron plasma with 
ultraintense fields.

As shown earlier in Refs.~\cite{efimenko.scirep.2018} and~\cite{efimenko.pre.2019},
in this configuration there are two main nonlinear regimes of interaction. In 
the first regime, which is realized when the laser power is less than 20 PW, 
thin electron-positron plasma sheets are formed as a result of the development 
of a current instability~\cite{efimenko.scirep.2018}. If 
the threshold power exceeds 20 PW, pinching of 
electron-positron plasma is possible as a result of current 
contraction
~\cite{efimenko.pre.2019}. In this paper in order to study the 
influence of resampling method we consider the interaction of an ideal dipole 
wave of total power $P=27$~PW with a 
low-density plasma target, which acts as a seed for the development of a
QED cascade. The choice of this value of power can be justified by two factors. 
First, the development of a QED cascade in such a configuration was discussed 
in 
detail in~\cite{efimenko.pre.2019}, and, second, such a problem statement allows 
studying the 
dynamics of the system at both (linear and nonlinear) stages of evolution.

In the numerical simulation, performed with the PICADOR PIC-code \cite{SURMIN2016204,10.1007/978-3-319-49956-7_25}, 
the interaction 
of an ideal 27 PW dipole wave with a plasma target with a 
3 $\mu$m diameter and a density of $10^{13}$ cm$^{-3}$ is simulated. The 
size of the simulation area is set to $4 \times 4 \times 4$~$\mu$m$^3$ with a grid size of $512 \times 
512 \times 512$, which, for a radiation wavelength of $\lambda$ = 0.9~$\mu$m, corresponds 
to a resolution of 115 points per wavelength. The time step is 0.015 fs, which 
corresponds to 200 steps per period of laser radiation. This resolution is 
sufficient to resolve the dynamics of the electron-positron plasma. In this section we employ the previously discussed methods to resample 
the particle ensemble using either thinning or merging techniques. All methods 
except \textit{mergeAv} are compared among each other. The resampling 
coefficient $k=2$ is used in all schemes.
In papers ~\cite{efimenko.scirep.2018} and~\cite{efimenko.pre.2019}, the 
\textit{globalLev} method is used. For completeness of presentation, this 
method is also compared with other methods discussed in this paper. 
The simulations are carried out on the Cascade Lake section of the MVS-10P supercomputer of the Joint Supercomputer Center of RAS. We employ 11 computational nodes (512 cores overall), 1 MPI process per core. The 3D simulation area is decomposed into 8 computational domains in each direction.

\begin{figure}[t!]
	\centering
	\includegraphics[width=0.49\columnwidth]{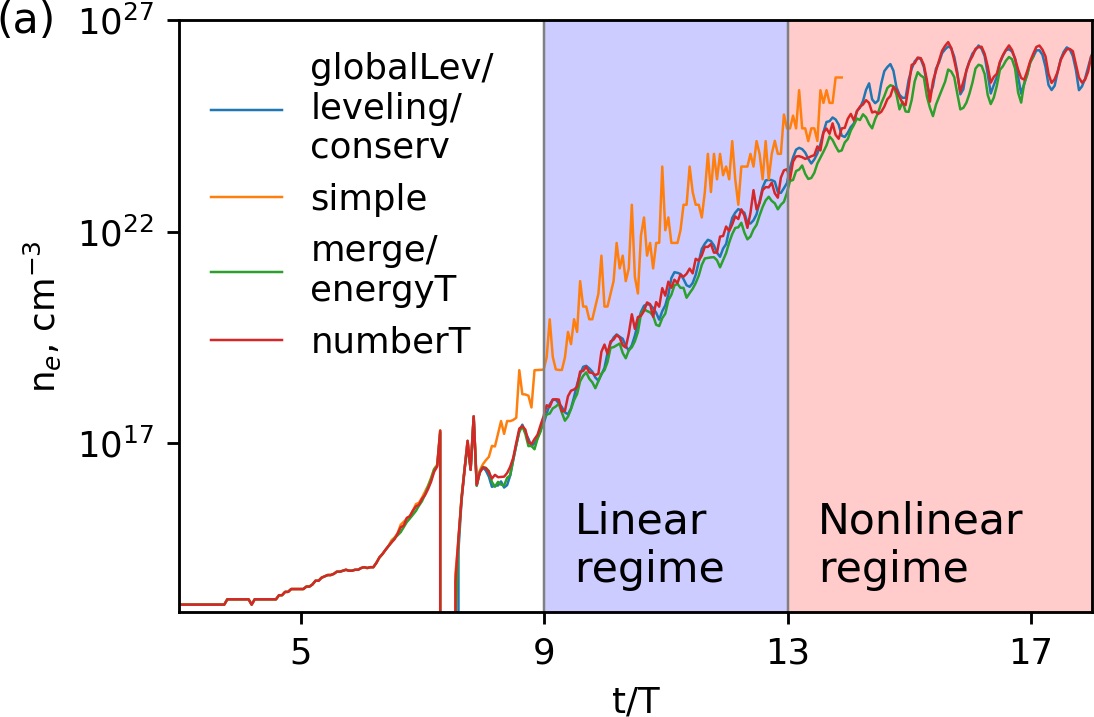}
	\includegraphics[width=0.49\columnwidth]{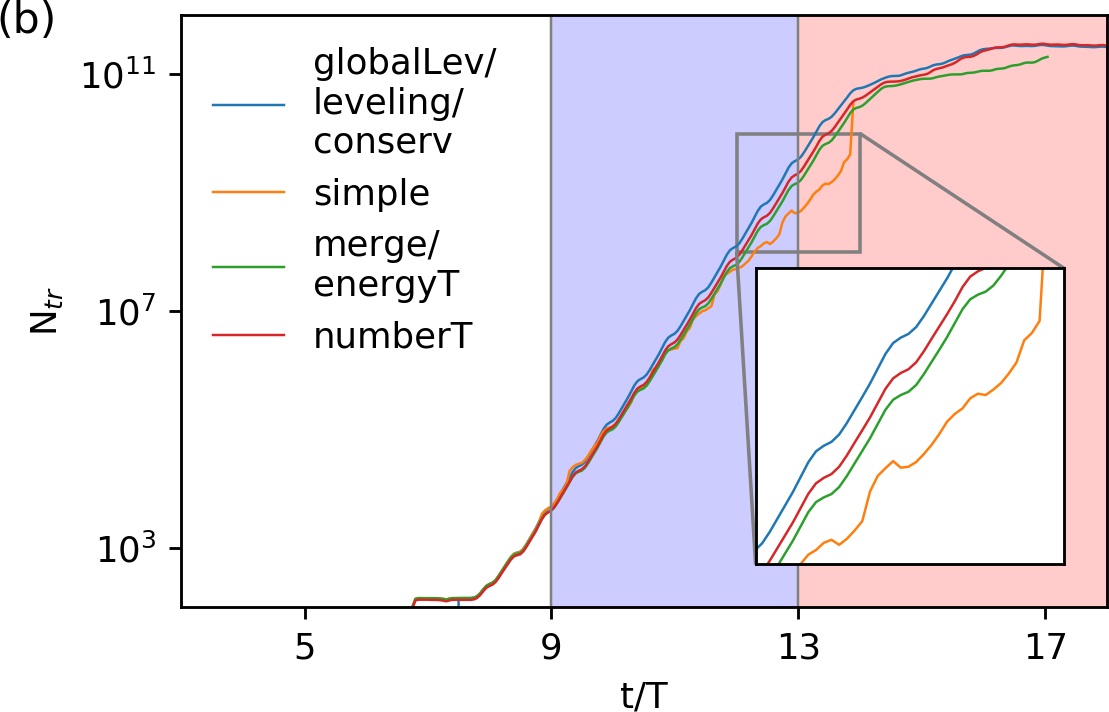}
	\caption{Temporal evolution of QED cascade in the field of a 
converging 27 PW dipole wave. (a) The 
maximum electron-positron plasma density in the $z=0$ plane during the development of the QED cascade. (b) The total number of electrons (positrons) in the cylinder with a 
diameter and height equal to $\lambda$.  
Methods with similar results are grouped accordingly. Color online.}
	\label{pinch}
\end{figure}

During the interaction the system evolves through several stages thoroughly 
discussed in related papers. First, the target is compressed towards the center 
of symmetry, which can be seen as the peak in maximum density at 7--8~$T$ in 
Fig.~\ref{pinch} (a). The next stage, if the initial target density is low enough, 
is a linear QED cascade in a given field. During this stage the maximum pair 
density and total number of particles grow exponentially. This stage is 
marked in blue in Fig.~\ref{pinch}. When pair plasma density becomes 
comparable with the critical density, the transition to the nonlinear regime 
occurs, which manifests itself in the form of saturation of the dependency of the plasma density
 (Fig.~\ref{pinch}(a)) and total particle number (Fig.~\ref{pinch}(b)) on time. The origin 
of such behaviour is high absorption in overdense electron-positron pair plasma 
which leads to a significant drop of the field amplitude. This stage is marked in red in Fig.~\ref{pinch}.

As noted previously, such a problem statement allows us 
to study both linear and nonlinear stages of the QED cascade. It should be 
expected that the choice of the resampling algorithm should have the most 
significant impact on the dynamics of the system in the linear regime, since 
upon transition to the nonlinear stage the growth rate significantly 
decreases and resampling of the ensemble of particles occurs less frequently.

\begin{figure}[t!]
	\centering
	\includegraphics[width=\columnwidth]{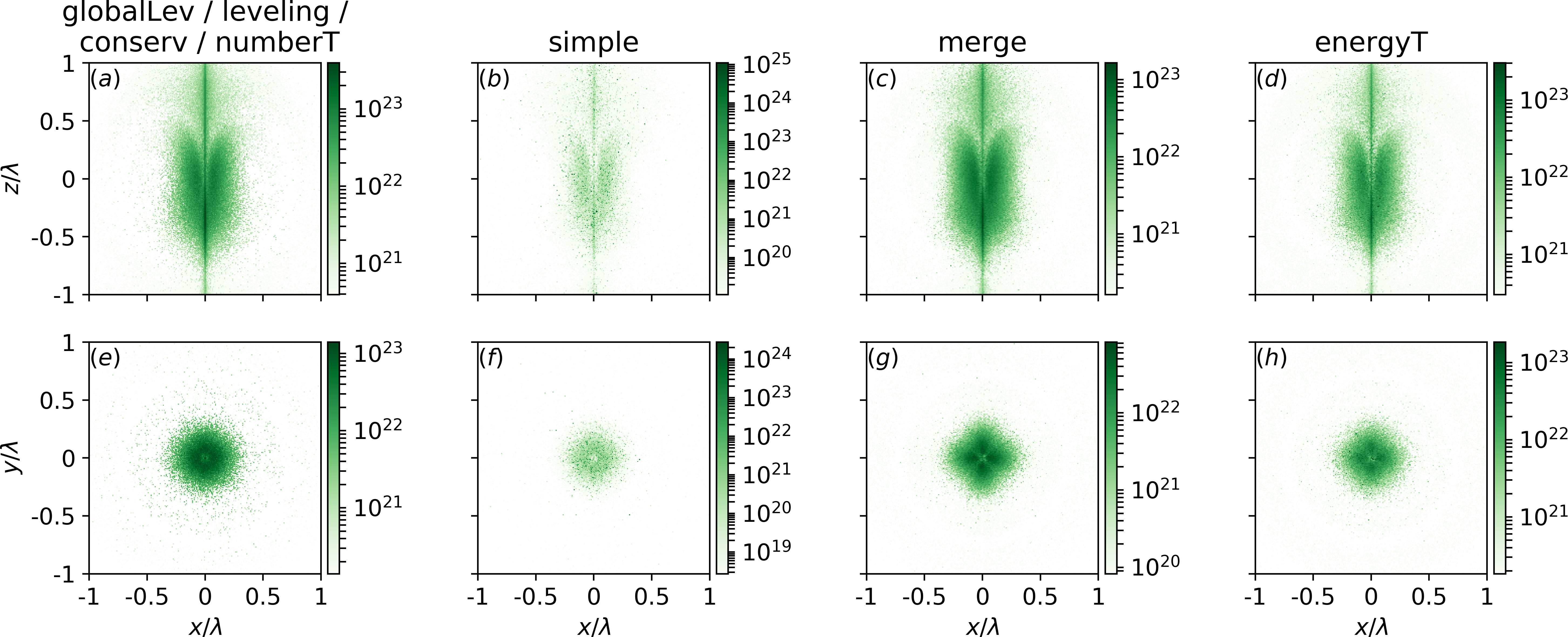}
	\caption{Spatial distribution of electron-positron plasma density in planes $x=0$ (a--d) and $z=0$ (e--h) in the end of linear 
stage of cascade ($t=13T$) for different resampling methods: 
(a, e) \textit{leveling, globalLev, conserv, numberT};  (b, f) \textit{simple};
(c, g) \textit{merge}; (d, h) \textit{energyT}. Note that color mapping is set individually for each subplot. Color online.}
	\label{pinch1}
\end{figure}

\begin{figure}[t!]
	\centering
	
\includegraphics[width=0.49\columnwidth]{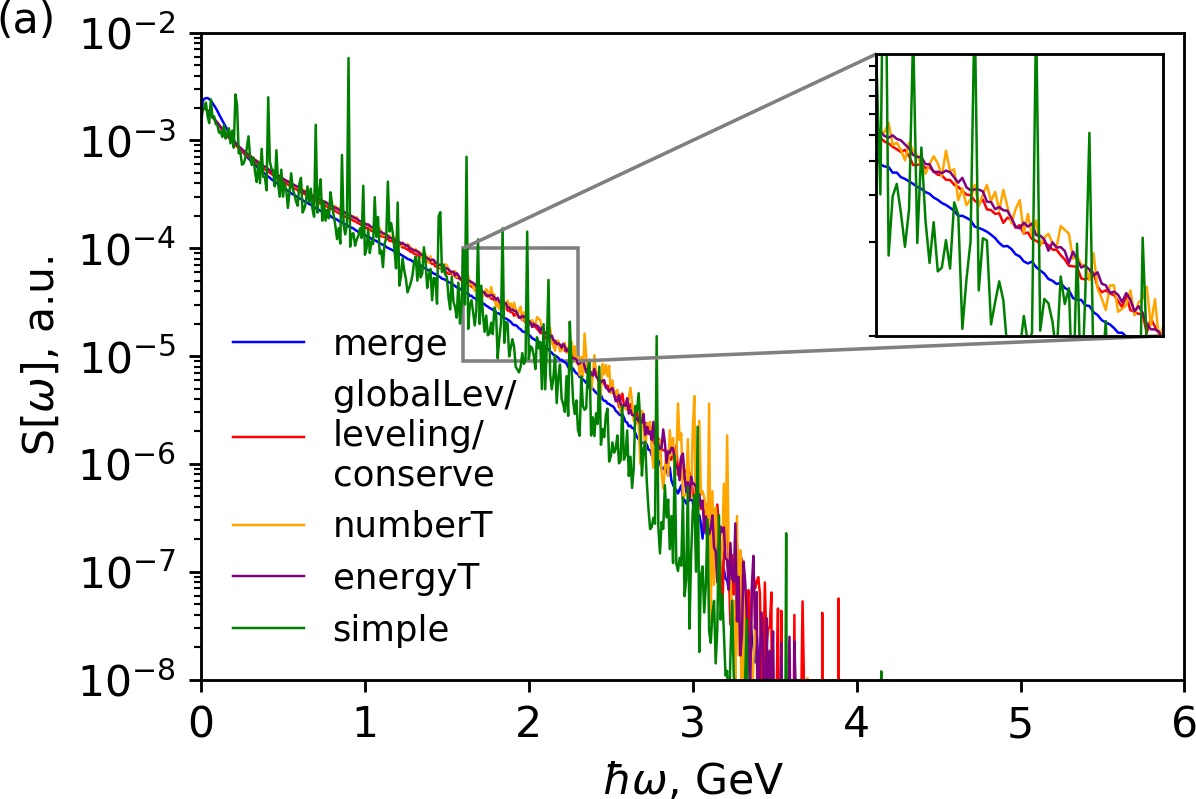}
	\includegraphics[width=0.49\columnwidth]{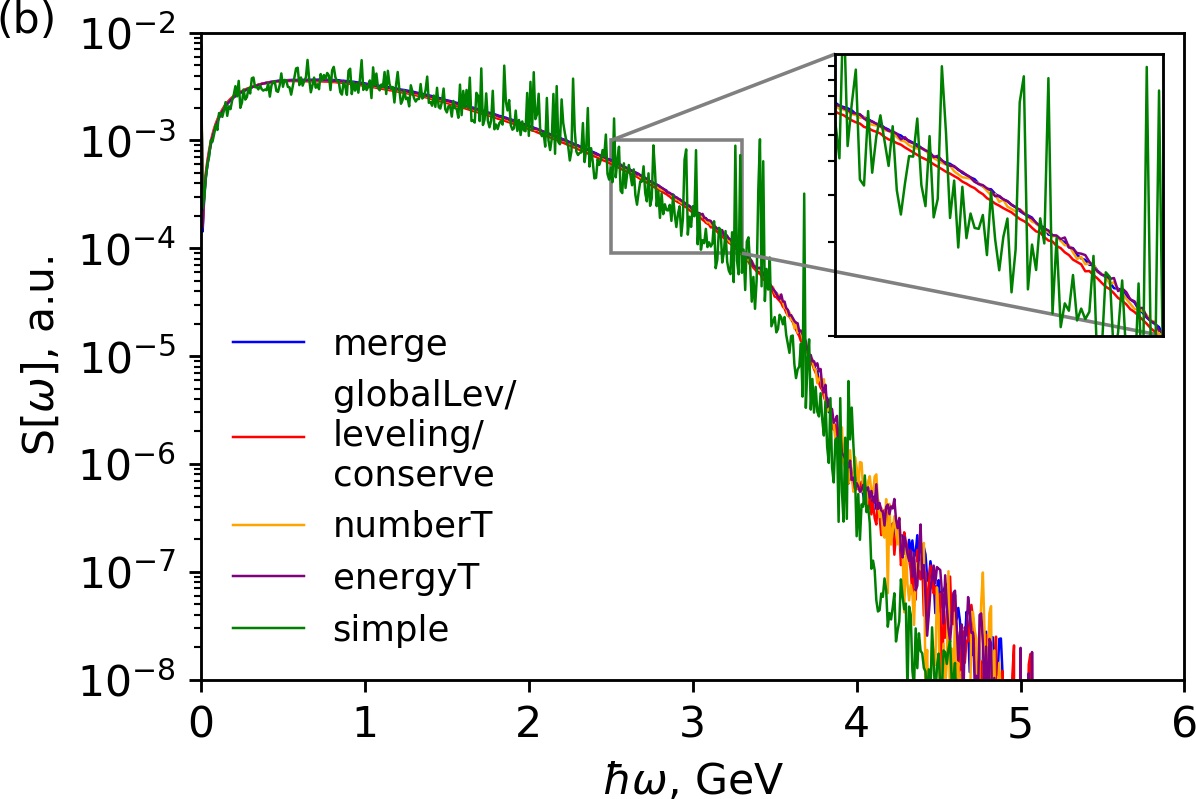}
	\caption{Energy spectra of (a) photons and (b) electrons 
(positrons) averaged over the linear stage ($t \in [9T, 13T]$) with different resampling methods. Methods with similar results are grouped 
accordingly. Color online.}

	\label{linear_spectra}
\end{figure}

Let us now compare and assess the physical results given by different resampling methods during the linear stage. Apart from the temporal evolution of the total particle number and peak density shown in Fig.~\ref{pinch}, we plot the spatial distribution of density in the end of the linear stage ($t = 13T$) in Fig.~\ref{pinch1}, average energy spectra of photons and electrons in Fig.~\ref{linear_spectra}, and also provide some essential data in Table~\ref{tablePinch}, including the measured value of the growth rate $\Gamma T$. According to these results, the resampling methods can be divided into several groups as noted in Table~\ref{tablePinch}.
The methods \textit{globalLev}, \textit{leveling} and \textit{conserv} behave almost identically, both at the linear and the nonlinear stages of development of the QED cascade. The growth rates yielded by these methods differ from each other by about $10^{-3}$. In Sec 5.1 these methods showed the closest results to those of the benchmark simulation without resampling. In this problem due to a high growth rate it is impossible to complete such a simulation. Given that the underlying algorithm behind \textit{conserv} is very different from that of \textit{globalLev} and \textit{leveling}, we can assume these methods cross-validate each other and consider this group of methods a benchmark.

In terms of the value of $\Gamma T$, the least deviation from the results of this benchmark is achieved by \textit{numberT} $(\sim3\%)$, a slightly larger deviation is shown by the group of \textit{merge} and \textit{energyT} $(\sim5\%)$ and the worst result is shown by \textit{simple} $(\sim10\%)$, as noted in Table~\ref{tablePinch}. Similarly, for the total number of particles over time (Fig.~\ref{pinch}(b)) the least deviation from the benchmark methods is achieved by \textit{numberT}, a slightly larger one by \textit{merge} and \textit{energyT}, and \textit{simple} performs the worst.

Fig.~\ref{pinch}(a) shows the maximum density of the electron-positron plasma during the development of the QED cascade. It is clear that for the method \textit{simple} the maximum value of pair plasma density exceeds the value obtained for all other methods and the distribution is exceptionally noisy. At the same time, the total number of particles is almost identical to other methods, which indicates that the problem is the formation of extraordinarily hefty macroparticles, which will be discussed below.

At the linear stage pair plasma evolves in the given field of the dipole wave, so the effects related to resampling are isolated from the self-consistent dynamics observed at the nonlinear stage. All resampling methods under such conditions produce similar plasma distributions, shown in Fig.~\ref{pinch1}. As discussed earlier, the methods \textit{globalLev}, \textit{leveling} and \textit{conserv} yield a similar symmetric distribution, which can be treated as the benchmark. Despite the notable difference in the cascade growth rate, the \textit{numberT} method yields a similar symmetric distribution and, therefore, all four methods are grouped together in Fig.~\ref{pinch1}(a,e).

Using the \textit{simple} resampling method leads to the formation of a large number of extraordinarily hefty macroparticles. Locally, these "super-particles" produce pair density two orders of magnitude higher than observed with other methods, see Fig.~\ref{pinch1}(b,f). Surprisingly, methods \textit{merge} and \textit{energyT} lead to the appearance of an azimuthal modulation of plasma density (see Fig.~\ref{pinch1}(c,g) and Fig.~\ref{pinch1}(d,h)), which is absent when other methods are used. Since these modulations are observed already at the linear stage of interaction, it seems that it is somehow caused by resampling, but an explanation for this effect requires a separate investigation.

\begin{table}[h!]
  \begin{center}
    \caption{Comparison of resampling methods for the problem of pinching of
electron-positron plasma in a 27 PW dipole wave. Run time, number of resampling 
instances, and average number of particles computed using a time interval of 100 
iterations, at linear and nonlinear stages, are given for the central process. Simulation of the nonlinear regime for method
\textit{simple} failed due to a numerical instability.}
    \label{tablePinch}
   
    \begin{tabular}{c|c|c|c|c|c|c|c}
    \hline
\multicolumn{2}{c|}{} & 
\multicolumn{3}{c|}{\textbf{Linear regime}} & 
\multicolumn{3}{c}{\textbf{Nonlinear regime}}\\
     \hline
    \textbf{Type} & \textbf{$\Gamma T$} & \thead{Time,\\ sec} & \thead{\# of \\ 
resampling} & \thead{\# of \\ particles \\ $\times 10^6$} & \thead{Time,\\ sec} 
& \thead{\# 
of \\ resampling} & \thead{\# of \\ 
particles \\ $\times 10^6$}  \\
      \hline
      \textit{globalLev} & 3.38 & 1771 & 28 & 3.11 & 734 & 6 & 3.07 \\
      \textit{leveling} & 3.38 & 1956 & 23 & 3.6 & 869 & 7 & 2.96 \\
      \textit{conserv} & 3.38 & 3744 & 65 & 4.45 & 1965 & 23 & 3.22 \\
      \hline
      \textit{numberT} & 3.27 & 2181 & 6 & 4.2 & 1100 & 7 & 2.6 \\
      \hline
      \textit{merge} & 3.19 & 3219 & 36 & 4.73 & 23983 & 23 & 4.3 \\
      \textit{energyT} & 3.21 & 1874 & 25 & 5.1 & 830 & 9 & 3.17 \\
      \hline
      \textit{simple} & 3.07 & 2373 & 31 & 5.1 & - & - & - \\
      \hline
   \end{tabular}
  \end{center}
\end{table}

We provide the spectra of photons and electrons(positrons) in Fig.~\ref{linear_spectra}. Comparison of spectra for different methods shows that at the linear stage of the QED cascade all methods yield similar results, showing that in the case of a large number of particles the distribution functions are preserved on average to a good degree. The \textit{simple} method due to formation of hefty particles yields a very non-uniform spectra with high peaks related to a high weight of these particles. The \textit{leveling}, \textit{globalLev} and 
\textit{conserv} methods lead to nearly indistinguishable spectra, although the \textit{conserv} and \textit{globalLev} methods have less noise at high frequencies. The \textit{merge} method has a photon energy distribution slightly biased towards low energy particles.

\begin{figure}[t!]
	\centering
\includegraphics[width=0.49\columnwidth]{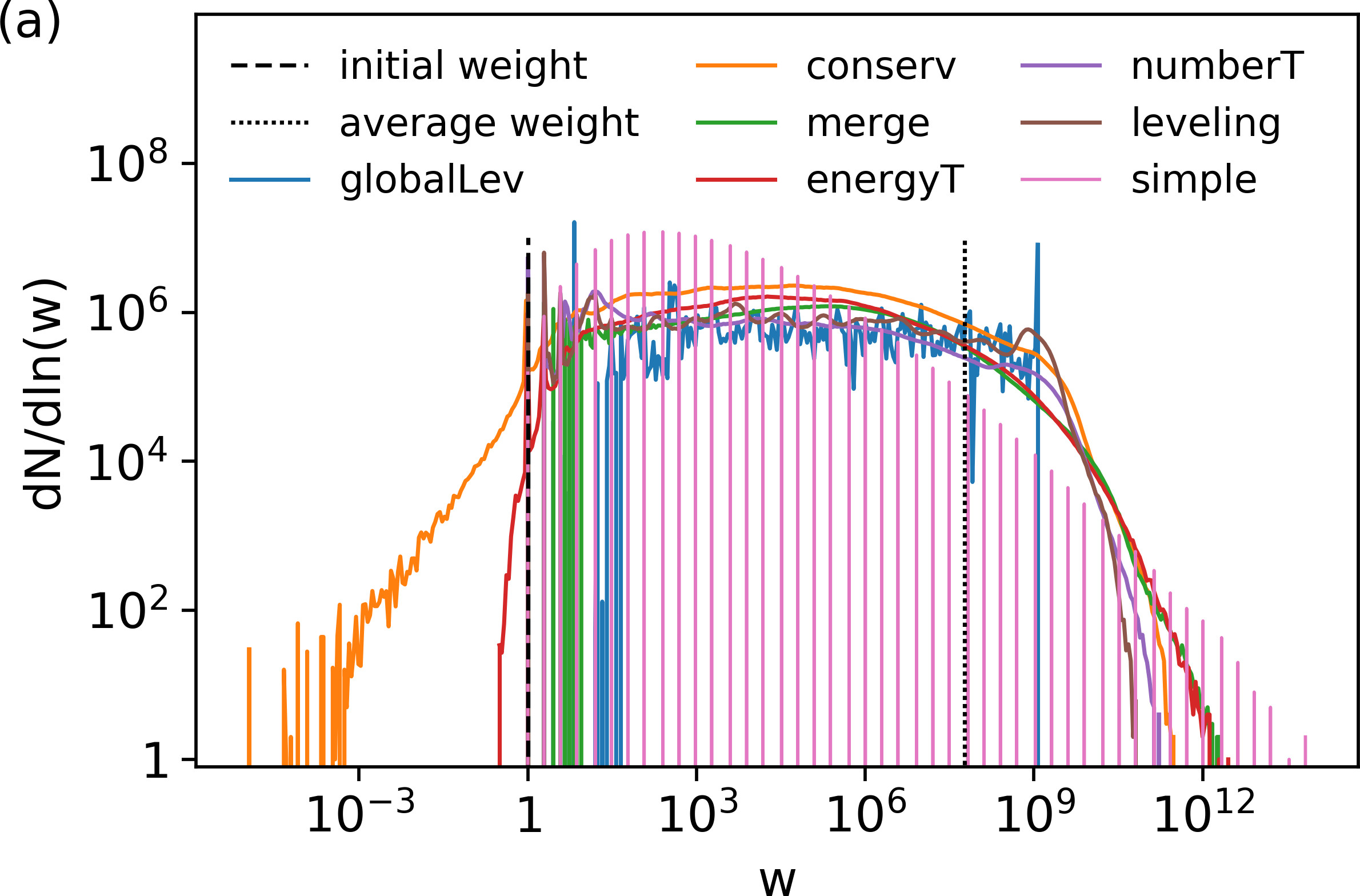}
	\includegraphics[width=0.49\columnwidth]{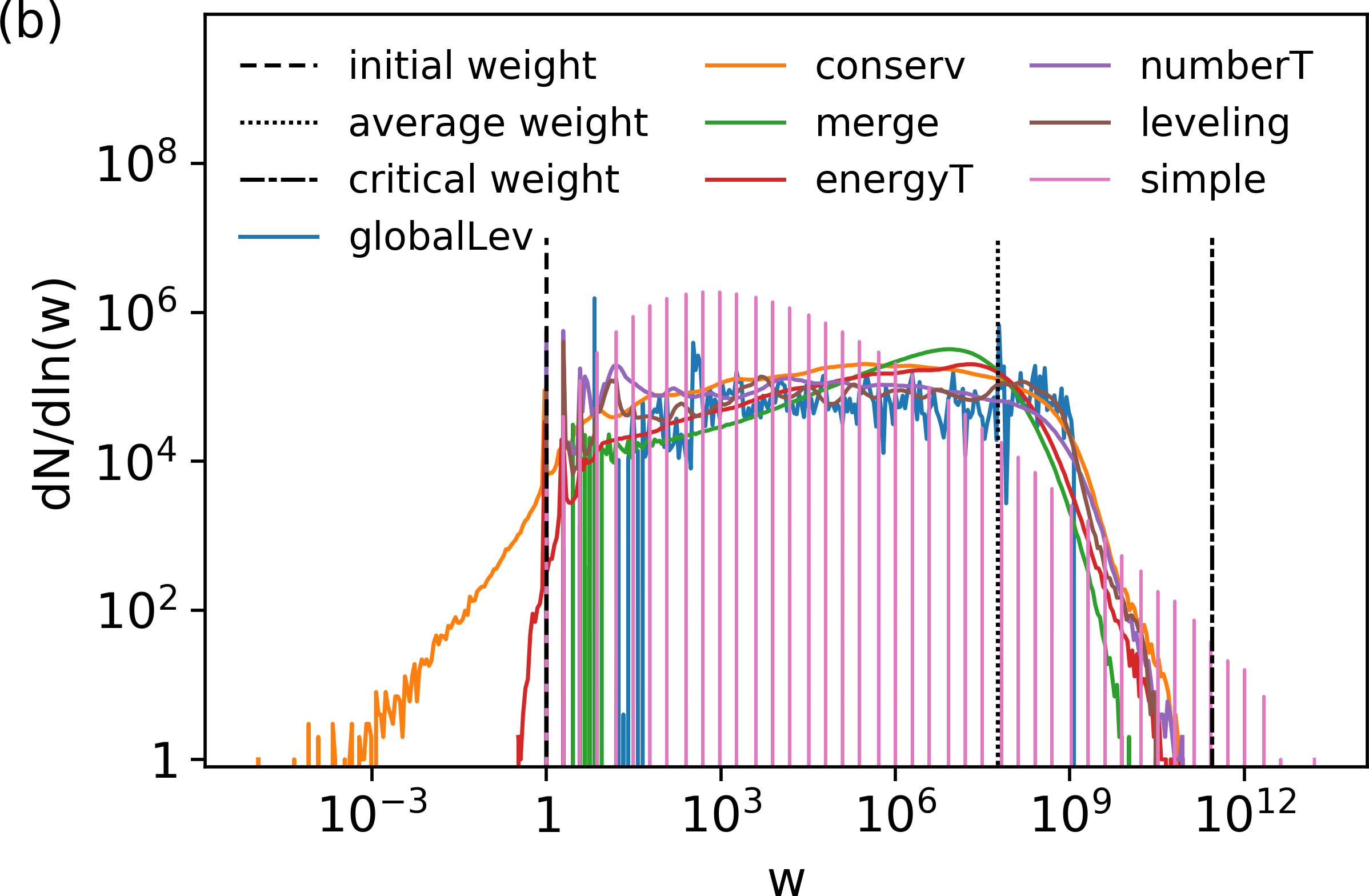}
	\caption{The distribution of weights in the end of the linear stage ($t=13T$) computed separately for macroparticles that sample photons (a) and electrons/positrons (b). The results are given for different resampling methods, which are indicated by the colors of the curves. All the weights are normalized to initial weight shown by the dashed line. The dashed-dotted line shows the critical weight ($\sim 2.75 \times 10^{11}$), single particle with this weight creates relativistic critical density in a cell. On both panels dotted line depicts the average electron weight for the \textit{globalLev} method ($\sim 5.7 \times 10^{7}$). The width of the bin for weight $w$ is 0.1$w$. 
	Color online.}
	\label{factors}
\end{figure}

We would also like to point out a rather straightforward factor that is likely to play a large or even dominant role in the interpretation of results obtained using resampling. The descriptive capability of particle ensembles primarily depends on the number of macroparticles and this is what restricts us from decreasing their number to reduce the computational costs. Nevertheless, the descriptive capability can be deteriorated if the weights of macroparticles start to differ largely. For example, if an ensemble is comprised of macroparticles of two largely different values of weight, only the hefty macroparticles effectively contribute to the descriptive capability (assuming that small features are not sampled intentionally by “light” macroparticles or such correlation is not maintained). From this ultimate case we see that an overly broad distribution of weights can lead to the degradation of descriptive capability. This can cause statistical noise and eventually affect the physical processes being simulated. To analyze the results in this respect we plot the distribution of macroparticles on their weights in Fig.~\ref{factors} and provide some hard data in Table~\ref{tableWeights}.

First of all, we would like to comment on the sets of possible values for particle weights given by different resampling methods. We note that for the sake of this study particles are initiated with the same weight. QED processes, the only source of new particles in the simulation other than resampling, create particles with weight always equal to that of the parent particle, so on its own these processes do not result in new values of particle weight.

\begin{table}[t!]
  \begin{center}
    \caption{Comparison of the distribution of weights computed for different resampling methods in the end of the linear stage ($t=13T$), as shown in Fig.~\ref{factors}. Minimum and maximum weights are normalized to the initial weight. Average weight and standard deviation are normalized to average electron weight for the \textit{globalLev} method ($\sim 5.7 \times 10^{7}$).}
    \label{tableWeights}
    \begin{tabular}{c|c|c|c|c|c|c}
    \hline
\multicolumn{1}{c|}{} & 
\multicolumn{2}{c|}{\textbf{\renewcommand{\arraystretch}{0.75}\begin{tabular}{@{}c@{}} Minimum weight/ \\ Maximum weight\end{tabular}}} &
\multicolumn{2}{c|}{\textbf{Average weight}} & 
\multicolumn{2}{c}{\textbf{Std deviation}}\\
     \hline
    \textbf{Type} & \thead{Electron} &  \thead{Photon} & \thead{Electron} & 
\thead{Photon} & \thead{Electron} & \thead{Photon}  \\
\hline
\textit{globalLev} &
\renewcommand{\arraystretch}{0.5} \begin{tabular}{@{}c@{}} 1 / \\ $1.2 \times 10^{9}$ \end{tabular}  &  \renewcommand{\arraystretch}{0.5} \begin{tabular}{@{}c@{}} 1 / \\ $1.2 \times 10^{9}$ \end{tabular} & 1  &  1.81  &  2.78  &  5.37  \\
\hline
\textit{leveling} &  \renewcommand{\arraystretch}{0.5}\begin{tabular}{@{}c@{}} 1 / \\ $3.0 \times 10^{10}$ \end{tabular}  &  \renewcommand{\arraystretch}{0.5}\begin{tabular}{@{}c@{}} 1 / \\ $6.4 \times 10^{10}$ \end{tabular}  & 0.92  &  1.46  &  3.46  &  7.56  \\
\textit{conserv} & 
\renewcommand{\arraystretch}{0.5}\begin{tabular}{@{}c@{}} $1.1 \times 10^{-5}$ / \\ $8.7 \times 10^{10}$ \end{tabular}  &  \renewcommand{\arraystretch}{0.5}\begin{tabular}{@{}c@{}} $1.1 \times 10^{-5}$ / \\ $3 \times 10^{11}$ \end{tabular}  & 0.57  &  0.72  &  3.63  &  9.03  \\
\textit{numberT} &  \renewcommand{\arraystretch}{0.5}\begin{tabular}{@{}c@{}} 1 / \\ $8.5 \times 10^{10}$ \end{tabular}  &  \renewcommand{\arraystretch}{0.5}\begin{tabular}{@{}c@{}} 1 / \\ $1.7 \times 10^{11}$ \end{tabular}  &  0.43  &  0.76  &  3.17  &  7.54  \\
\hline
\renewcommand{\arraystretch}{0.5}\textit{merge} &  \renewcommand{\arraystretch}{0.5}\begin{tabular}{@{}c@{}} 1 / \\ $3 \times 10^{10}$ \end{tabular}  &  \renewcommand{\arraystretch}{0.5}\begin{tabular}{@{}c@{}} 1 / \\ $2\times 10^{12}$ \end{tabular}  & 0.27  &  0.64  &  1.1  & 
 19.89  \\
\textit{energyT} &  \renewcommand{\arraystretch}{0.5}\begin{tabular}{@{}c@{}} 0.33 / \\ $8.0 \times 10^{10}$ \end{tabular}  &  \renewcommand{\arraystretch}{0.5}\begin{tabular}{@{}c@{}} 0.31 / \\ $2.8\times 10^{12}$ \end{tabular}  & 0.39  &  0.51  &  1.97  &  
17.47  \\
\hline
\textit{simple} & 
\renewcommand{\arraystretch}{0.5}
\begin{tabular}{@{}c@{}} 1 / \\ $1.7 \times 10^{13}$ \end{tabular} & 
\renewcommand{\arraystretch}{0.5}
\begin{tabular}{@{}c@{}} 1 / \\ $6.7 \times 10^{13}$ \end{tabular} & 0.12  &  0.13  & 
 77.79  &  
184.3  \\
\hline
  \end{tabular}
  \end{center}
\end{table}

As evident from its mechanics, the method \textit{simple} produces a discrete set of possible values of particle weights: $w_i=k^i$, where $k$ is the resampling parameter defined in Sec.3 and $i= 0, 1, 2, ...$. All other methods except \textit{globalLev} produce weights individual to each particular cell, so integration over the whole computational domain yields an effectively continuous distribution. Global leveling, however, uses the same value $k \bar{w}$, the minimal possible weight after resampling, for all particles of the same type in the whole simulation. In the absence of QED effects and given that initially all particles have the same weight, the \textit{globalLev} method behaves identically with the \textit{simple} method. QED effects, however, result in newly-born particles inheriting the weight of their parent particles (of a different type). Together with resampling being performed for particles of different types independently (i.e. in the general case on different iterations), this means that every instance of resampling spawns a single new possible value for the particles’ weight equal to $k \bar{w}$ at that moment in time for particles of the respective type. As a result, the number of possible values for the particle weight is limited by the number of instances of resampling, but this number can be large enough to justify calling the \textit{globalLev} distribution of particle weights quasi-continuous.

With this in mind we analyze the weight distributions given in Fig.~\ref{factors}(a) (for photons) and Fig.~\ref{factors}(b) (for electrons) presented by histograms with log-scale bins, meaning each bin represents the number of macroparticles with weights between $f$ and $rf$, where $f$ is the current bin's minimal weight and $r>1$ is the bin's width on the logarithmic scale. We choose $r=1.1$, which allows us to emphasize the \textit{simple} distribution's discrete nature. The particles' weight is normalized to the initial weight. The dash-dotted line illustrates the critical weight which enables a single macroparticle to represent within its cell a density equal to the characteristic relativistic critical density $\gamma N_c = \frac{a_0}{2} N_c$, where $\gamma$ is the maximal Lorentz-factor of particles, $a_0 \simeq 4100$ is the amplitude of the electric field in relativistic units for a 27 PW e-dipole wave, $N_c = 1.34 \times 10^{21}$ cm$^{-3}$ is the critical density for the wavelength $\lambda = 0.9$ $\mu m$. Such a particle alone is capable of significantly affecting the electromagnetic field. For given parameters the critical weight is equal to $2.75 \times 10^{11}$. We note that this value depends on the cell size employed in the simulation. Since the distribution is presented at the end of the linear regime (during which plasma density is supposedly insufficient to affect the field), one might argue that no single macroparticle should be able to affect the field. Therefore we can conclude that the \textit{simple} method, yielding 58 times larger weights leads to incorrect results.

The method that stands out most according to the distributions in Figure~\ref{factors} is the \textit{globalLev} method. Its maximal weight is fewer than that of other methods by 1-2 orders of magnitude for both electrons and photons, which provides for a smoother representation of the particle ensemble.

As one can see, the \textit{simple} method provides the broadest distribution of weights. The fact that this method leads to the highest level of noise among all the methods (see Fig.~\ref{pinch1}) can therefore be associated with the related deterioration of the descriptive capability of the ensemble. The high level of noise also follows from the fact that for the \textit{simple} method the peak density notably exceeds that of other methods (see Fig.~\ref{pinch}(a)), while the total number of electrons(positrons) does not deviate that remarkably (see Fig.~\ref{pinch}(b)). Presumably, in this case we see how an overly high level of noise affects the physical processes and eventually alters the value of growth rate. As discussed in Sec 5.2 and as can be seen from Figure~\ref{factors}, the probabilistic nature of the \textit{simple} method may (and does) lead to overly large maximal particle factors compared to other methods, and, as a result, to an abnormally large physical particle density in some cells, which may lead to an unphysical instability.

Let us now consider the possibility of a similar (but weaker) effect when employing other methods of resampling. The methods \textit{leveling} and \textit{globalLev} are specifically designed to affect only particles with a low weight, and the weight of the resulting particles is limited by $k \bar{w}$, which expectedly mitigates this effect. The methods \textit{numberT}, \textit{conserv}, \textit{merge}, and \textit{mergeAv} strictly conserve the number of physical particles in a cell, so it cannot artificially increase due to these methods of resampling. For the method \textit{energyT} there is no such restriction, so the total amount of physical particles in a cell may potentially increase significantly, especially because of low energy particles whose weight increases inverse-proportionally to energy (this is balanced on average by the principle of agnostic resampling). However, in comparison with the \textit{simple method}, the conservation of energy makes it less likely, and in our experience the development of artificial instabilities with the \textit{energyT} method has not been observed.

Note that an overly large number of physical particles in a cell is the criterion for the development of the particular numerical instability described in the previous section that we have experienced throughout our work on simulation of interaction of plasmas with ultraintense laser fields. In physical problems of a different kind other characteristics could be more important, and resampling methods could be modified accordingly.

For convenience we present parameters of weight distributions, shown in Fig.~\ref{factors}, in Table~\ref{tableWeights}. Minimal and maximal weights are normalized to the initial weight and show the range of weights given by a certain resampling method, while the average weight and its standard deviation characterize the distribution as a whole. To facilitate the comparison of these two last values for different resampling methods, we scale them to the average electron weight yielded by the \textit{globalLev} method, depicted on both panels in Fig.~\ref{factors} by the dotted line. One may wonder whether the difference in average macroparticle weight reflects the difference in physical results or in the weight distributions. As evident from Fig.~\ref{pinch} and Fig.~\ref{pinch1}, physical quantities (such as the number of physical particles and even their spatial distribution) given by the methods \textit{globalLev}, \textit{leveling}, and \textit{conserv} can be considered equal, thus the discrepancy in average weights for these methods can be fully attributed to the difference in the number of macroparticles. For other methods the number of physical particles differs significantly for the purposes of this particular argument, so the discrepancy in average weights is the result of the combination of the two causes.

As before, the methods with similar properties are grouped together. According to the previous discussion the \textit{globalLev} method is grouped separately. It has the narrowest range of weights and the greatest average weight among other methods. The \textit{leveling, conserv} and \textit{numberT} methods have close properties, but the maximal weight is at least an order of magnitude higher than that given by the \textit{globalLev} method. The standard deviation is also higher, but is quite similar for electrons and photons. The \textit{merge} and \textit{energyT} methods result in maximal electron(positron) weights close to the ones given by other methods, but the maximal weights for photons are two orders of magnitude higher, which indicates that with these methods photons with large weights do not produce any pairs. The last group, as before, is the \textit{simple} method which yields a 4 order of magnitude higher maximal weight as compared to the \textit{globalLev} method, and a much higher standard deviation.

\begin{figure}[t!]
	\centering
	\includegraphics[width=0.75\columnwidth]{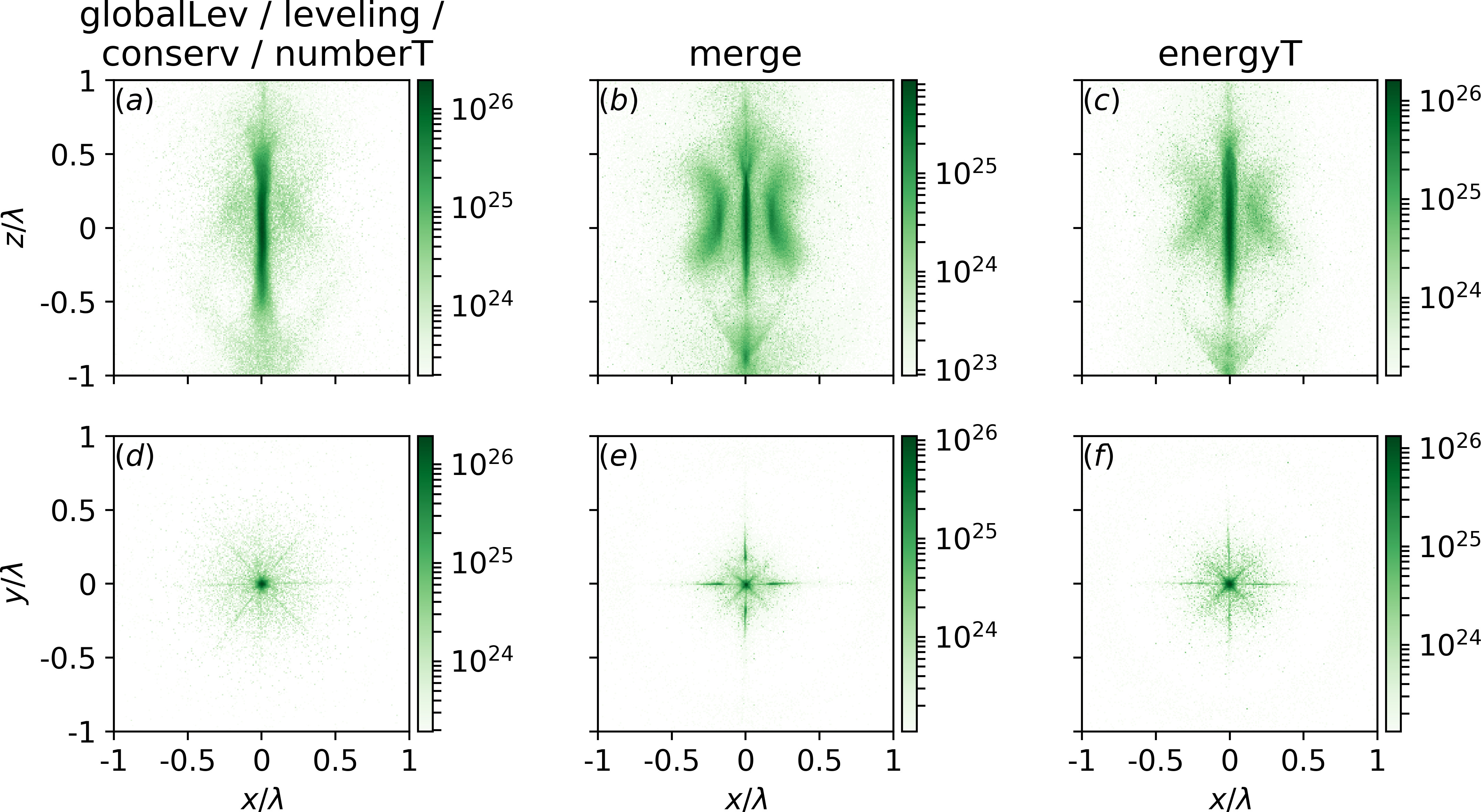}
	\caption{Spatial distribution of electron-positron plasma density in planes (a--c) $x=0$ and (d--f) $z=0$ at 
nonlinear stage of cascade ($t=16.6T$) 
for different resampling methods: 
(a, d) \textit{globalLev, leveling,conserv,numberT}; 
(b,e) \textit{merge}; (c, f) \textit{energyT}. The 
density of electron-positron plasma is plotted to a logarithmic scale. Color online.}
	\label{pinch2}
\end{figure}

At the nonlinear stage the analysis of different methods becomes non-trivial, 
because in this case dynamics becomes self-consistent and even slight deviations in 
pair density may be significantly enhanced. Nevertheless, for completeness of 
presentation we show some results from the nonlinear stage. First of all, when 
the \textit{simple} method is used the presence of hefty macroparticles leads 
to the formation of local density peaks and an earlier transition to the nonlinear 
regime compared to other methods, seen as a saturation of a total number of 
particles in Fig.~\ref{pinch}(b). The presence of hefty macroparticles and the 
consequent development of a numerical instability results in the termination of 
the numerical simulation at the beginning of the nonlinear stage. In this regime method grouping similar to the one observed at linear stage is retained. All other methods capture the main physical effect studied here --- the pinching of electron-positron pair plasma~\cite{efimenko.pre.2019}. 
The \textit{leveling},
\textit{globalLev}, \textit{conserv} and \textit{numberT} methods cause a formation of a slight azimuthal modulation with some of the peaks along quite random directions, that can 
be related to the development of a current instability~\cite{efimenko.scirep.2018} 
(see Fig.~\ref{pinch2}(a,d) ), i.e. physical in nature.
If the \textit{merge} or the \textit{energyT} method is used, the spatial modulation 
observed at the linear stage is enhanced, and this results in a formation of a cross-like transverse distribution, see 
Fig.~\ref{pinch2}(b,e) and (c,f). 

\begin{figure}[t!]
	\centering
\includegraphics[width=0.49\columnwidth]{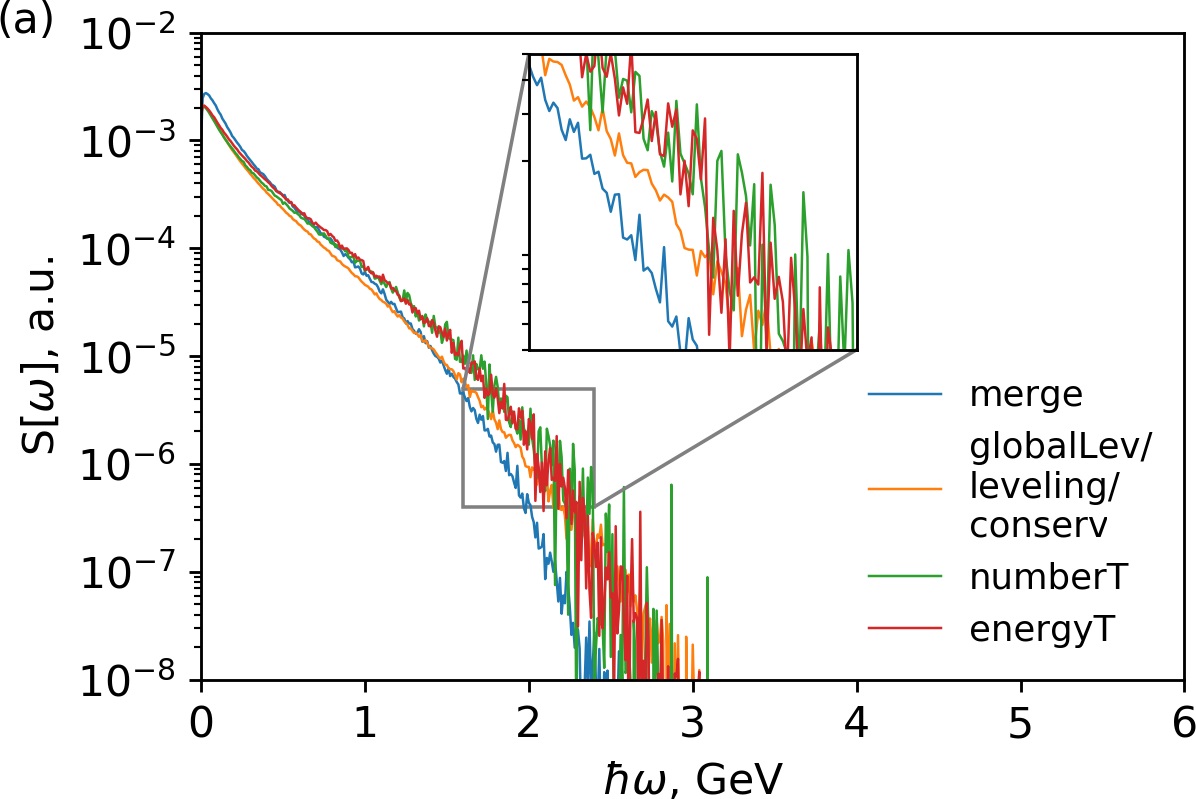}
	\includegraphics[width=0.49\columnwidth]{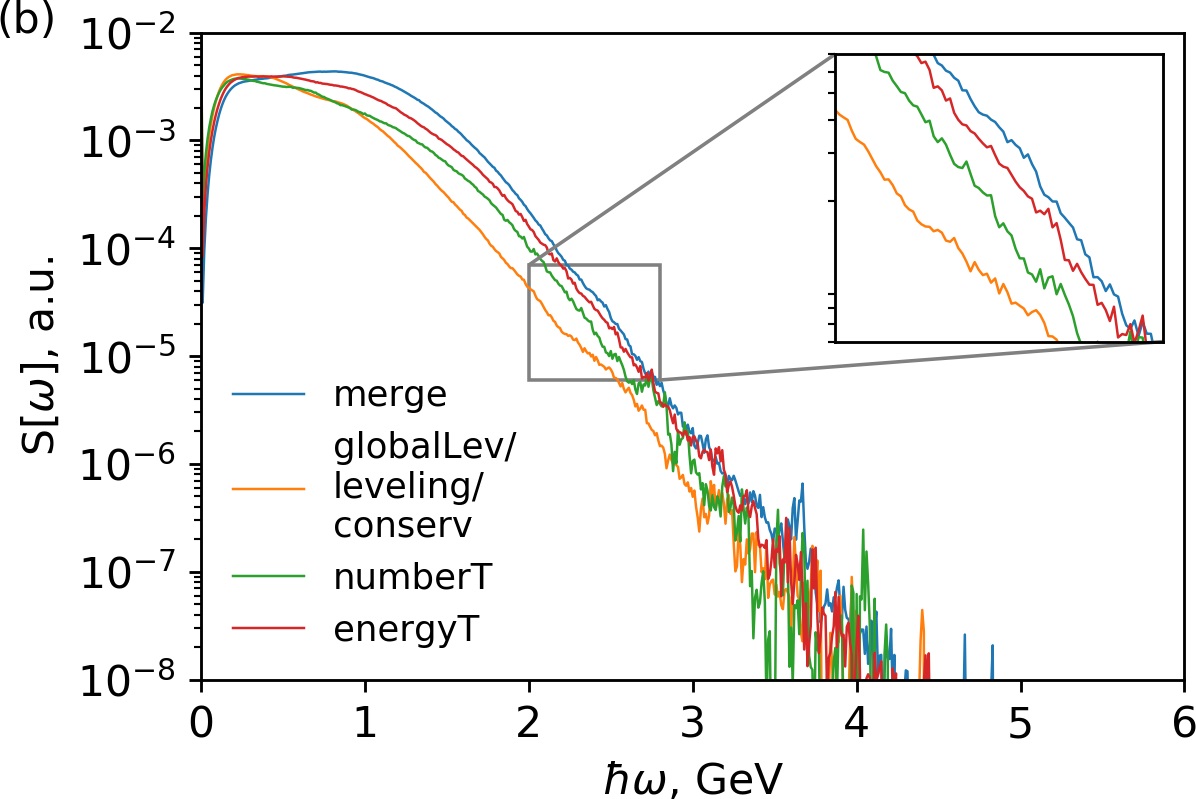}
	\caption{Energy spectra of (a) photons and (b) electrons 
(positrons) averaged over the nonlinear stage ($t \in [15T, 17T]$) with different resampling methods. Methods with similar results are grouped 
accordingly. Color online.}
	\label{nonlinear_spectra}
\end{figure}

The difference in spectra is more pronounced at the nonlinear regime, see 
Fig.~\ref{nonlinear_spectra}, but similar conclusions can be made. 
The photon spectra for the \textit{merge} method are biased in favor of low energy 
particles, while the \textit{energyT} and \textit{numberT} methods have slightly 
higher values in the middle part of the spectrum, see 
Fig.~\ref{nonlinear_spectra}(a). At the same time all methods except 
\textit{merge} have similar behavior near the maximum energy. For electron 
(positron) spectra the difference is mainly in the middle part, which 
can be also attributed to a slightly different self-consistent dynamics at 
the nonlinear stage of interaction. 

The simulation details for different methods are summarized in Table~\ref{tablePinch}. It can be seen that the simulation time, the number of times resampling is performed and the average number of particles significantly differ for different methods. At the nonlinear stage of the cascade the number of runs of resampling drops in comparison with the linear stage, thus the total running time also decreases. All thinning methods show similar performance, the \textit{conserv} method due to higher computational complexity has an approximately twofold running time. Surprisingly, the \textit{merge} method works quite well at the linear stage, but at the nonlinear stage the running time becomes prohibitive.    

Summarizing all aspects of the studied problem, we can conclude that all methods 
except \textit{simple} performed reasonably well, but the three 
best methods both at linear and nonlinear stages are the \textit{globalLev}, \textit{leveling} and \textit{conserv} methods, among which \textit{globalLev} is more computationally efficient.

\section{Conclusion}

The principle of agnostic down-sampling that is applicable without any prior knowledge about the problem was formulated and several resampling methods complying with this principle were presented. Results acquired with use of these methods were compared among each other, and also to theoretical results, results acquired without resampling (where possible) and to the results acquired by some non-agnostic methods. The comparison was performed first using simple model problems, and then using pertinent problems involving generation of plasma via QED cascade and thus often requiring extensive resampling.

It was shown that the relative accuracy of various methods highly depends on the problem at hand and the criteria for determining this accuracy.  Therefore we conclude that there is no universal method of resampling which would show the best performance in all cases. However, it can be noted that several methods provide stable performance on all problems that we have considered. These methods are \textit{leveling}, \textit{globalLev}, \textit{conserv}, and to a lesser extent \textit{energyT} and \textit{numberT}. The methods \textit{simple} and \textit{merge/mergeAv} have at least one example where the method in question significantly alters the physical outcome even though in certain other conditions these methods might be the most advantageous. It should also be noted that merge-based methods \textit{merge/mergeAv} significantly increase the computation time in comparison to the thinning methods. Most of the considered methods are released as open-source tools within the hi-$\chi$ framework \cite{hi-chi}.

\section{Acknowledgements}

The authors would like to acknowledge several funding sources for supporting different parts of the work, as specified in parenthesis: the Ministry of Education and Science of the Russian Federation, contract No. 14.W03.31.0032 being executed at the Institute of Applied Physics of the Russian Academy of Sciences (the simulation of laser-plasma interactions and the nonlinear QED cascades, Sec. 5); the Russian Foundation for Basic Research and the government of the Nizhny Novgorod region of the Russian Federation, grant No.~18-47-520001 (the development and comparison of the resampling methods); the Russian Foundation for Basic Research, grant No.~20-21-00095 (the development and optimization of algorithms for numerical modeling of QED cascades). A.M. acknowledges the Support of the "BASIS" Foundation grant No.~19-1-5-94-1 (theoretical basis for modeling of current sheets with QED cascades, Sec. 5.2). A.G. acknowledges the support of the Swedish Research Council, grant No.~2017-05148. The authors acknowledge the use of computational resources provided by the Joint Supercomputer Center of the Russian Academy of Sciences and by the Swedish National Infrastructure for Computing (SNIC).

\bibliography{literature}

\end{document}